\documentclass{aa} 
\usepackage{graphicx}
\usepackage{txfonts}
\usepackage{natbib}
\bibpunct{(}{)}{;}{a}{}{,}  

\mathchardef\period=\mathcode`.
\DeclareMathSymbol{.}{\mathord}{letters}{"3B}

\def\kpc{{h^{-1} \rm kpc}}
\def\kms{{\rm km\, s^{-1}}}
\def\aap{{\em A}\&{\em A}}

\def\apjl{{\em ApJ}}
\def\apjs{{\em ApJS}}

\begin{document}

   \titlerunning{Properties of galaxies with ring structures}
   \authorrunning{Fernandez et al. 2021}

   \title{Properties of galaxies with ring structures}


   \author{Julia Fernandez\inst{1},
          Sol Alonso\inst{1},
                  Valeria Mesa\inst{2,3},
                  Fernanda Duplancic\inst{1} \and Georgina Coldwell\inst{1}
          }

   \institute{Departamento de Geof\'{i}sica y Astronom\'{i}a, CONICET, Facultad de Ciencias Exactas, F\'{i}sicas y Naturales, Universidad Nacional de San Juan, Av. Ignacio de la Roza 590 (O), J5402DCS, Rivadavia, San Juan, Argentina\\
              \email{fernandezmjulia22@gmail.com }
         \and  
Departamento de Astronom\'{i}a, Facultad de Ciencias, Universidad de La Serena, Av. Juan Cisternas 1200 Norte, La Serena, Chile
\and 
Instituto Argentino de Nivolog\'{i}a Glaciolog\'{i}a y Ciencias Ambientales (IANIGLA-CONICET), Parque Gral San Mart\'{i}n, CC 330, CP 5500, Mendoza, Argentina\\
             }
             
   \date{Received xxx; accepted xxx}

   \abstract
   {}
   {We present a statistical analysis of different characteristics of ringed spiral galaxies with the aim of assessing the effects of rings on disk galaxy properties.
   }
   {We built a catalog of ringed galaxies from the Sloan Digital Sky Survey Data Release 14 (SDSS-DR14). Via visual inspection of SDSS images, we classified the face-on spiral galaxies brighter than  $g < 16 \period 0$ mag into galaxies with: an inner ring, an outer ring, a nuclear ring, both an inner and an outer ring, and a pseudo-ring. In addition to rings, we recorded morphological types and the existence of bars, lenses, and galaxy pair companions with or without interaction. With the goal of providing an appropriate quantification of the influence of  rings on galaxy properties, we also constructed a suitable control sample of non-ringed galaxies with similar redshift, magnitude, morphology, and local density environment distributions to those of ringed ones.}
   {We found 1868 ringed galaxies, accounting for  22\% of the full sample of spiral galaxies. In addition, within galaxies with ringed structures, 46\% have an inner ring, 10\% an outer ring, 20\% both an inner and an outer ring, 6\% a nuclear ring, and 18\% a partial ring. Moreover, 64\% of the ringed galaxies present bars. 
We also found that ringed galaxies have both a lower efficiency of star formation activity and older stellar populations (as derived with the $D_n(4000)$ spectral index) with respect to non-ringed disk objects from the control sample.  Moreover, there is a significant excess of ringed galaxies with red colors. 
These effects are more important for ringed galaxies that have inner rings and bars with respect to their counterparts that have some other types of rings and are non-barred. 
The color-magnitude and color-color diagrams show that ringed galaxies are mostly concentrated in the red region, while non-ringed spiral objects are more extended to the blue zone.
Galaxies with ringed structures present an excess
of high metallicity values compared to non-ringed ones, which show a $12 + \rm Log (\rm O /\rm H)$ distribution toward lower values. 
These findings seem to indicate that rings are peculiar structures that produce an accelerating galactic evolution, strongly  altering  the  physical properties of their host galaxies.
} 
   {}


\maketitle 
%

\section{Introduction}
 
Galaxy features have a significant impact on the understanding of how galaxies form and evolve. Internal and external processes, such as bar and spiral torques, mergers, tidal effects, starbursts, gas accretion, etc., occur over cosmic time, serving to modify galaxies. As such, the properties, structure, and morphology of galaxies represent meaningful approaches for tracing these processes \citep{Alonso2012, Lambas2003, Lambas2012, Mesa2014, Ho2019, Vollmer2013, Sanchez2014}.

Rings are galactic structures that have a specific relevance to the mechanisms involved in the evolution of galaxies. Rings are elliptical or circular stellar and gaseous features whose distinctive character can be intimately connected to specific aspects of galaxy dynamics.
In the local Universe, about 
one-fifth of all spiral galaxies include a ring-shaped pattern in their light distributions, 
and an additional third appear to have partial rings \citep[also known as pseudo-rings;][]{Buta1996}.

Different mechanisms appear to be involved in ring formation. For instance, \cite{Buta1996} make a clear distinction between what they called ``catastrophic rings," resulting from collisions between galaxies,  and ``normal resonant rings," created from accumulations of gas at certain resonances due to the continuous action of gravity torques produced by the bars.

Galaxy collisions can result in three types of rings: accretion rings (formed by accreted satellite galaxies;  \citealt{Schweizer1987}), polar rings (where a small, gas-rich galaxy is disrupted along a near polar orbit around a more massive disk galaxy, usually
an S0 galaxy), and collisional rings or R galaxies (formed when a small galaxy collides with the rotation axis of a larger disk galaxy, causing a radially expanding density wave; \citealt{Lynds1976}). Thus, there is a close relationship between the formation rate of collisional ring galaxies and the galaxy interaction rate \citep{Donghia2008}.

On the other hand, resonant rings make up the majority of observable ring galaxies, and the main resonances produce three types of rings: (i) nuclear rings, associated with the inner Lindblad resonance (ILR), (ii) inner rings, associated with the inner 4:1 resonance (I4R), and (iii) outer rings, associated with the outer Lindblad resonance \citep[OLR;][]{Buta2017}. 

Regarding how galaxy properties are affected in the presence of rings, some authors have carried out studies of particular ringed galaxies. For example, \cite{Gusev2003} studied the structure and stellar population of the ringed barred galaxy \object{NGC 2336}, finding a star formation rate (SFR) typical of late-type spiral galaxies. They also found that the colors of the outer disk are typical of an old stellar system with a small fractional contribution from a young stellar population. 
In addition, \cite{Grouchy2010} examined the SFRs of two samples, one composed of 18 barred and the other 26 non-barred ringed galaxies. They found that both samples exhibit similar star-forming properties as the inner rings and are not dependent on the shapes of the rings or the strength of the bar.

Several authors have explored some of the physical properties behind ringed galaxies.  \cite{Silchenko2018} studied the formation of stars in two early-type galaxies with outer rings. Others have attempted to determine if there is a relationship between the direction of the winding of the spiral arms, the presence of a bar, and the formation of the rings, as well as their influence on star formation in a normal inner ring \citep{Buta2003, Grouchy2008}. All found that it is difficult to generalize about the physical properties behind ringed galaxies due to inconsistent and incomplete existing data.

At present, several catalogs and samples of ringed galaxies are available at different wavelengths and provide us with valuable information, including, the Catalogue of Southern Ringed Galaxies, (CSRG; \citealt{Buta1995}) and the GZ2 Catalogue of Northern Ringed Galaxies (GZ2-CNRG; \citealt{Buta2017}), which are optical catalogs designed to confront morphological predictions of test-particle models (\citealt{Schwarz1981,Schwarz1984}; \citealt{Byrd}). Other studies (e.g., \citealt{Buta2015} and \citealt{Comeron2014}) have examined the same characteristics using images obtained at near-infrared wavelengths. Statistically significant samples are required to successfully analyze the properties of ringed galaxies and thus provide an adequate description of the physics underlying these objects.

In this paper we have built a homogeneous catalog of ringed galaxies with the intent to examine how the properties of galaxies are affected in the presence of rings. In this way we attempt to expand this little-studied subject with a complete statistical sample. In addition,  this catalog will serve as a complement for future studies aimed at shedding light on this topic.

This paper is structured as follows. Section 2 describes the database of our catalog and the procedure used to construct the samples of ringed galaxies as well as the control sample selection criteria. In Sect. 3 we study the galaxy properties in different types of ringed galaxies, focusing the analysis on the SFRs, stellar populations, color indices, and metallicity in host spiral galaxies with respect to non-ringed ones. Finally, Sect. 4 summarizes our main conclusions. The adopted cosmology throughout this paper is $\Omega_m=0.3$, $\Omega_{\Lambda}=0.7$, and $H_0=70 \kms \rm Mpc ^{-1}$.


\section{Catalog of ringed galaxies}
\subsection{Sample selection}
This research is based on data selected from the Sloan Digital Sky Survey \citep[SDSS;][]{York2000} Data Release 14  \citep[DR14;][]{Abolfathi2018}. DR14 is the second data release of the fourth phase of the SDSS \citep[SDSS-IV, 2014-2020;][]{sdssiv}. It contains five types of data: images, optical spectra, infrared spectra, integral field unit spectra, and catalog data such as parameters measured from images and spectra. Furthermore, DR14 includes all previous versions. 

In the present work we consider spectroscopic and photometric data from the SDSS-DR14.
For this data set, k-corrected absolute magnitudes were estimated from Petrosian apparent magnitudes converted to the AB system. The k-corrections band-shifted to $z=0.1$ were calculated using the software \texttt{k-correct\_v4.2} of \cite{Blanton2007}. In our analysis we use the u, g and r-bands in the \textit{ugriz} system. In addition, we obtained several physical properties available in the SDSS spectroscopic database through Structured Query Language (SQL) in CasJobs\footnote{http://skyserver.sdss.org/casjobs/}. 
We extracted: gas-phase metallicities, stellar masses, specific
SFRs, concentration index, etc. \citep{Balogh1999, Brinchmann2004, Kauffmann2003, Tremonti2004}.

With the aim of obtaining ringed galaxies, we first considered restrictions in redshift, $ 0 \period 01 <z <0 \period 1 $, and magnitude $ g <16 \period 0 $ to exclude all those galaxies whose morphological details are difficult to detect by visual inspection. 
Furthermore, as was noted by \cite{Buta2017}, rings become harder to detect visually with increasing disk inclination. In particular, an inner ring could be difficult to detect at high inclination due to foreshortening and internal extinction.
Hence, we applied an additional constraint on the ellipticity of the objects, selecting galaxies with axial ratio $ b/a> 0 \period 5 $.

Finally, to discriminate between bulge and disk galaxy types, a  constraint was imposed on the concentration index C\footnote{$C=r90/r50$ is the ratio of Petrosian 90 \%- 50\% r-band light radii} \citep{Abraham1994}, a well tested morphological classification parameter  \citep{Strateva2001},  also used as a good stellar-mass tracer ($M_*$) and an indirect index of the SFR \citep{Deng2013}.  \citet{Yamauchi2005} performed a galaxy morphological classification using the C parameter, finding very good agreement with the visual classification.
Furthermore, in order to obtain late-type galaxies we selected objects with a concentration index value $ C <2 \period 8 $.
In this way, bright spiral-type galaxies with face-on orientation  and $ 0 \period 01 <z <0 \period 1 $ were selected, thus favoring the subsequent visual classification process. 
With these constraints, our sample comprises 8529 galaxies, and, therefore, we can make a plausible visual inspection of a large set of objects.

\subsection{Classification}

\begin{figure}
\centering
\includegraphics[width=0.4\textwidth]{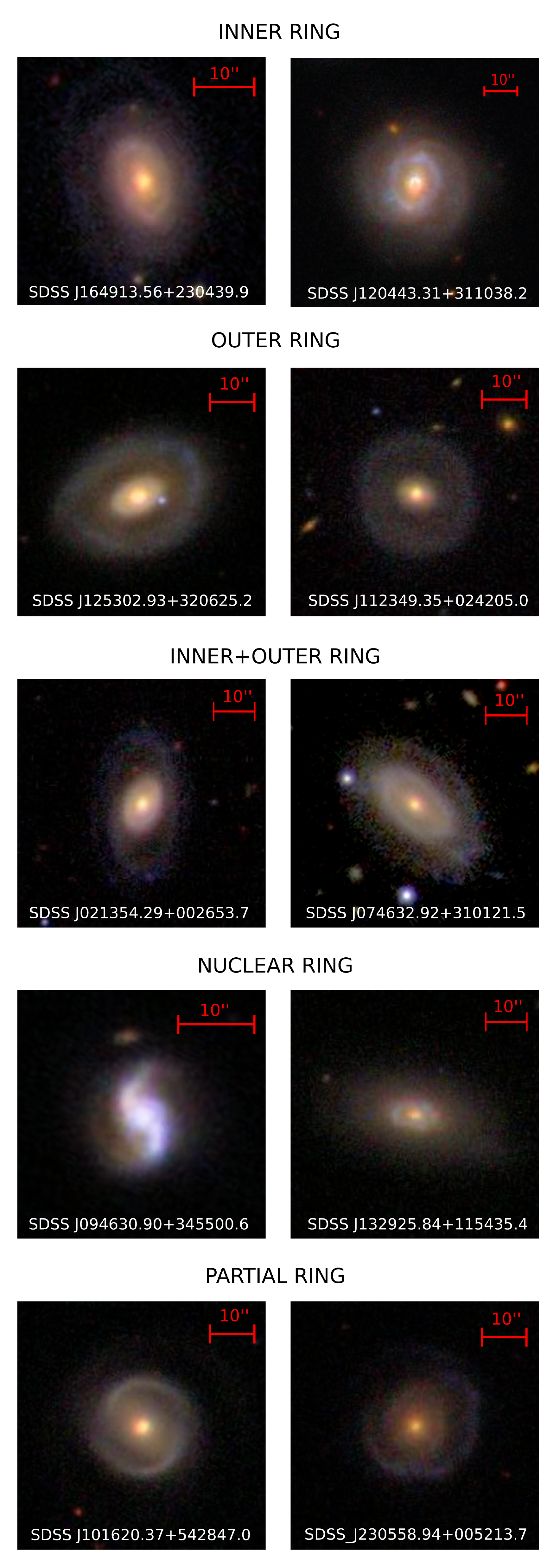} 
\caption{Examples of the different types of rings considered. 
}
\label{fig:rings} 
\end{figure}

In this section we describe the ringed galaxies selection process by visual inspection. For this task, we used the g + r + i combined color images, obtained from online SDSS-DR14 Image List\footnote{https://skyserver.sdss.org/dr14/en/tools/chart/listinfo.aspx}. Then, via a thorough visual examination, we classified the galaxies based on the presence of ringed structures. Rings were divided into five types: inner, outer, both inner and outer, nuclear, and partial, as shown in Fig. \ref{fig:rings}. 
Consequently, if the galaxy presents a ringed structure, other features were also taken into account, such as: morphological types, bars, lens and pair companions with and without interaction evidence.

With regards to the morphological classification we adopted the system that represents the Hubble stage of galaxies from elliptical to irregular: S0$^-$/S0/S0$^+$, S0/a, Sa, Sab, Sb, Sbc, Sc, Scd, Sd, Sdm, Sm, and Im \citep[e.g.,][]{Nair2010,Buta2015}. 
In this system we can distinguish between S0$^-$, S0 and S0$^+$ where the superscript ``-'' denotes ``early'' and the ``+'' denotes ``late.'' Galaxies with doubtful morphology were flagged with a question mark. A more detailed description of the different morphological types is available in \cite{Buta2015}.

\begin{table}
\center
\caption{Numbers and percentages of galaxies with different types of rings (see Fig. \ref{fig:rings}).}
\begin{tabular}{|c c c| }
\hline
Ring Type & Number &  Percentage   \\
\hline
\hline
\textsc{inner ring} & 857 & 46\% $\pm$ 3.9\% \\
\textsc{outer ring} &  186 & 10\% $\pm$ 0.7\%  \\
\textsc{inner + outer rings} & 372 & 20\% $\pm$ 2.1\%     \\
\textsc{nuclear ring} &  111 & 6\% $\pm$ 0.8\%  \\
\textsc{partial ring} &  342 & 18\% $\pm$ 1.9\%  \\
\hline
TOTAL  & 1868 & 100\% \\ 
\hline
\end{tabular}
{\small}
\label{tab:Table1}
\end{table}

\begin{figure}[h!]
\centering
\includegraphics[width=0.5\textwidth]{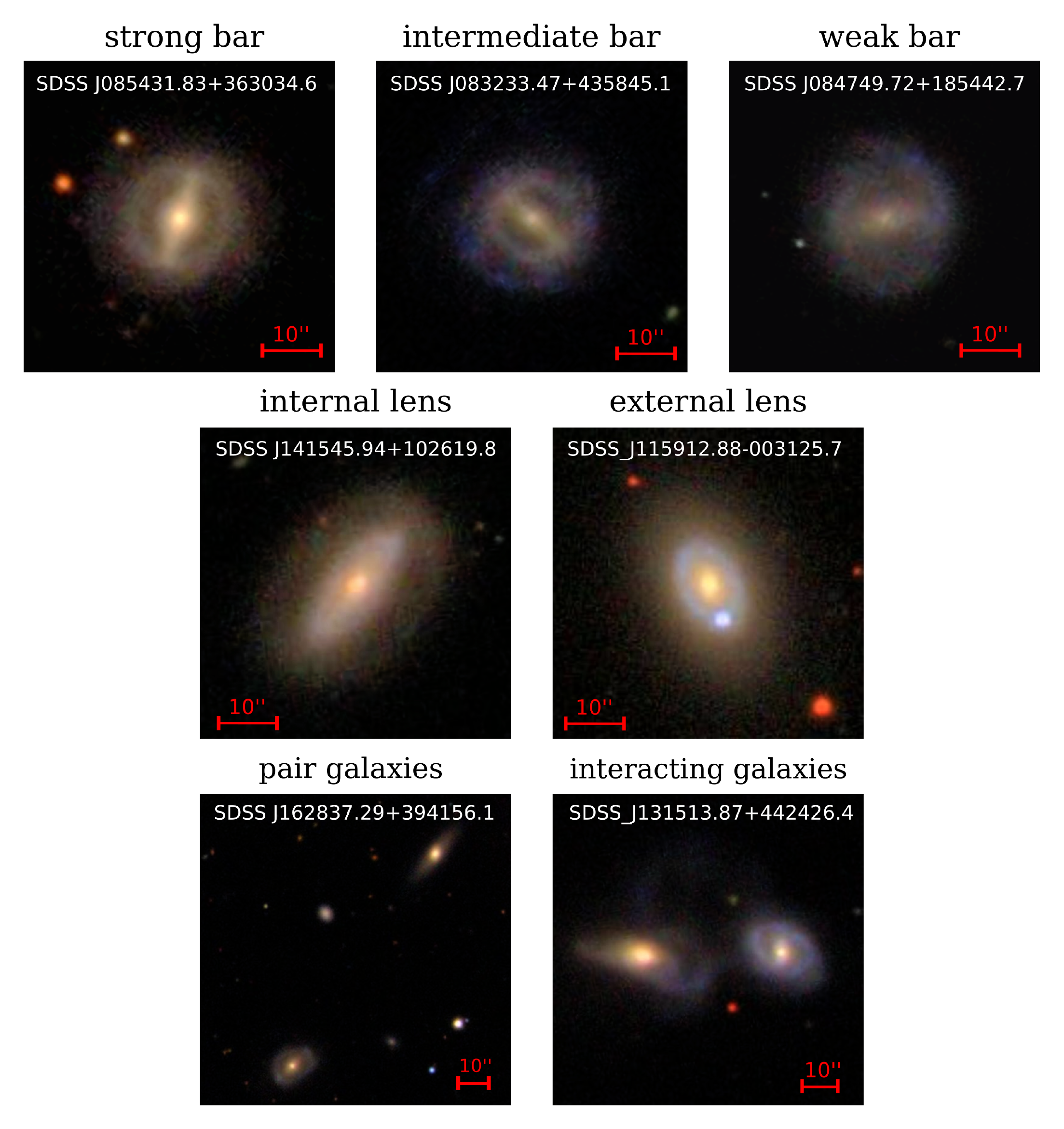} 
\caption[Different types of bars, lenses and ringed galaxies in pairs considered.] {Examples of the different types of bars (top panels), lenses (middle panels), and ringed galaxies in pairs (bottom panels) considered.}
{\label{fig:barras_lenses_pairs}}
\end{figure} 

In addition to the ring classifications, we also recognized the presence of strong, intermediate, and weak bars (see Fig. \ref{fig:barras_lenses_pairs}, upper panels), based on the criteria used by \cite{Buta2015} for SA,  S\underline{A}B, SAB, and SB galaxies.
Further, two types of lens were also taken into account: internal and external lens (see Fig. \ref{fig:barras_lenses_pairs}, middle panels). 
It should be noted that different combinations of lenses were not considered in the present work.

For completeness, we identified ringed galaxies in pair systems.
\cite{Lambas2003} and \cite{Alonso2006} found that projected  distance $r_p<100 \kpc$  and  difference radial  velocity $\Delta V<350 \kms$ are convenient thresholds for stellar formation activity induced by interactions. In particular for a sample of spiral galaxy pairs, \cite{Mesa2014} adopted $r_p<50 \kpc$  and $\Delta V<500 \kms$ values.
In this direction, we considered ringed galaxies  with  a  close  companion  within $r_ {p}<50h^{- 1} kpc$ and $\Delta V<350 \kms$. 
Then, the systems were classified according to two categories:
(i) pairs: ringed galaxies inhabiting pair systems that show no distortions in their morphologies and (ii) interactions: ringed galaxies in close pairs,  where the gravitational fields generate morphological disturbance.
The bottom panels of Fig. 2 show examples of these categories.

\begin{table*}[h!]
\centering
\caption{Numbers and percentages of galaxies with different types of rings and morphologies.}\label{tab:Table2}
\resizebox{\textwidth}{!}{
\begin{tabular}{|l|r|r|r|r|r|r|r|r|r|}\hline
Ring Type &S0$^-$/S0/S0$^+$&S0/a&Sa&Sab&Sb&Sbc&Sc&Sd&?\\
\hline
\hline
\textsc{inner ring} &34 (2\%)&105 (6\%)&395 (21\%)&147 (8\%)&126 (7\%)&6 (0.32\%)&32 (2\%)&1 (0.053\%)&11 (1\%)\\ \hline
\textsc{outer ring} &14 (1\%)&27 (1\%)&133 (7\%)&7 (0.37\%)&2 (0.11\%)&0 (0\%)&0 (0\%)&1 (0.053\%)&2 (0.11\%)\\ \hline
\textsc{inner + outer rings} &4 (0.21\%)&5 (0.26\%)&355 (19\%)&6 (0.32\%)&2 (0.11\%)&0 (0\%)&0 (0\%)&0 (0\%)&0 (0\%)\\ \hline
\textsc{nuclear ring} &21 (1\%)&51 (3\%)&21 (1\%)&4 (0.21\%)&7 (0.37\%)&0 (0\%)&2 (0.11\%)&0 (0\%)&5 (0.26\%) \\ \hline
\textsc{partial ring} &2 (0.11\%)&15 (1\%)&131 (7\%)&112 (6\%)&54 (3\%)&6 (0.32\%)&11 (1\%)&2 (0.11\%)&9 (0.48\%) \\  \hline
\hline
TOTAL & 75 (4\%) & 203 (11\%) & 1035 (55\%) & 276 (15\%) & 191 (10\%) & 12 (1\%) & 45 (2\%) & 4 (0.21\%) & 27 (1\%) \\ \hline
\end{tabular}}
\label{tab:Table2}
\end{table*}

\begin{table*}[h!]
\renewcommand{\tablename}{Table} 
\centering
\caption{Numbers and percentages of galaxies with different types of rings related to bars, lens, and pair systems (see Fig. \ref{fig:barras_lenses_pairs}).} 
\resizebox{\textwidth}{!}{
\begin{tabular}{|l|l|l|l|l|l|l|l|l|}
\hline
&\multicolumn{3}{l|}{Barred}&\multicolumn{2}{l|}{Lens}&\multicolumn{2}{l|}{Pairs}\\
\hline
\hline
 Ring Type&Strong&Intermediate&Weak&Inner&External&Interaction&No interaction\\
\hline
\hline
\textsc{inner ring} &220 (12\%)&266 (14\%)&190 (10\%)&71 (4\%)&132 (7\%)&21 (1\%)&50 (3\%)\\ \hline
\textsc{outer ring} &11 (1\%)&7 (0.37\%)&12 (1\%)&65 (3\%)&20 (1\%)&3 (0.16\%)&10 (0.53\%)\\ \hline
\textsc{inner + outer rings}
&59 (3\%)&111 (6\%)&102 (5\%)&88 (5\%)&7 (0.37\%)&2 (0.11\%)&20 (1\%)\\ \hline
\textsc{nuclear ring} &1 (0.053\%)&0 (0\%)&2 (0.11\%)&5 (0.3\%)&47 (3\%)&0 (0\%)&11 (1\%)\\ \hline
\textsc{partial ring} &40 (2\%)&102 (5\%)&76 (4\%)&21 (1\%)&2 (0.11\%)&7 (0.37\%)&30 (2\%)\\
\hline
\hline
TOTAL & 331 (18\%) & 486 (26\%) & 382 (20\%) & 250 (13\%) & 208 (11\%) & 33 (2\%) & 121 (8\%) \\ \hline
\end{tabular}}
\label{tab:Table3}
\end{table*}


Then, the adopted classification to build the ringed galaxy catalog is summarized as follows: 1868 ringed galaxies were obtained with respect to the full sample of spiral galaxies. This represents the 22\% of the total sample. 
We also found that from the total number of galaxies with ringed structures, 46\% present inner rings, 10\% outer rings, 20\% both inner and outer rings, 6\% nuclear rings, and 18\% partial rings. The visual inspection was performed by one of the authors in
order to maintain a unified criterion. The reliability of the classification was addressed by comparing it with the classification of a subsample of ringed galaxies from another author.
This procedure allows us to quantify the uncertainty in the classification for the different types of rings
(see Table \ref{tab:Table1}).

Taking into account the morphological types, we found 55\% are Sa galaxies, 15\% Sab, 11\% S0/a, 10\% Sb, 4\% S0$^-$/S0 /S0$^+$, 2\% Sc, 1\% Sbc, 1\% Doubtful, and 0.21\% Sd.
 Moreover, we found that 36\% of ringed galaxies are non-barred, 26\% present intermediate bars, 20\% weak bars and 18\% strong bars, respectively.
The percentages of ringed galaxies exhibiting lenses are: 13\% with internal and 11\% with external lens. 
Lastly, regarding pairs and interactions, it was found that 8\% are in pair systems and 2\% present interactions.
Tables  \ref{tab:Table2} and \ref{tab:Table3} show numbers and percentages related to the different considered features.
 As can be observed, most SDSS color images belong to the Sa morphological type, being those having an inner ring or an inner+outer combination the predominant ones. Most galaxies with bars, lenses and a companion are shown to be objects  with an inner ring, an intermediate bar and an external lens. In general 8\% of the total of galaxies in pairs show no evidence of interaction with the close companion.

\cite{Comeron2014} found that the relative frequency of outer rings and pseudo-rings decreases significantly for stages after Sb in the ARRAKIS (Atlas of Resonance Rings as known in the $S^{4}G$) catalog, while inner rings peaks at stages between S0$^+$ and Sb. According to \cite{Buta2017} this drop toward later stages may be due to late-type galaxies are under-represented in the catalog. This effect  is compound by the fact that inner rings are also the largest in size in early-type galaxies. Moreover, \cite{Buta2017} found that galaxies presenting weak bars are predominant in the ringed galaxies catalog obtained from the Galaxy Zoo 2 database \citep{Lintott2008}.
In agreement with our classification, \cite{Nair2010} found that the galaxies with an inner ring occur in 44 \% of the total sample considered, while 13 \% of the galaxies have an outer ring, 8 \% have both an inner and an outer ring, and 11 \% have a nuclear ring.
Some of our results are consistent with previous research, and the main differences may be due to the distinct classification criteria used in each case.

\section{Analysis of the galaxy properties}

In the previous section we described the construction of a ringed galaxy catalog with a basic classification aimed to answer important questions about the properties of ringed galaxies, as well as their nature, formation, and evolution.

In particular, in this paper we are concerned to explore the effect of the different types of rings on the star formation activity, stellar population, color and metallicity of the host galaxies, in comparison with the suitable control sample. 
Then, with the purpose to obtain reliable statistical  results, and taking into account the main morphological features, the ringed galaxy catalog was divided into four specific groups to carry out the subsequent analysis: (i) ringed galaxies with an inner ring (857 objects), (ii) ringed galaxies, including the remaining types: double rings (inner and outer), outer rings, nuclear rings, and partial rings (1011 objects), (iii) ringed barred galaxies with strong, intermediate, or weak bars (1199 objects), and (iv) ringed non-barred galaxies (669 objects).

 In the case of galaxies with inner ring and other types of rings, both groups may or may not present bars. In the same way, barred or non-barred ringed galaxies can exhibit any kind of ring. Furthermore, ringed galaxies in these four specific groups may or may not present lens, galaxy pair companions, and display different Hubble morphological classifications. 
 
 The detailed classification of the catalog, described in Sect. 2 will be useful for studies of every particular subsample in the upcoming papers.

 \subsection{Control sample}

\begin{figure}
\includegraphics[width=0.47\textwidth]{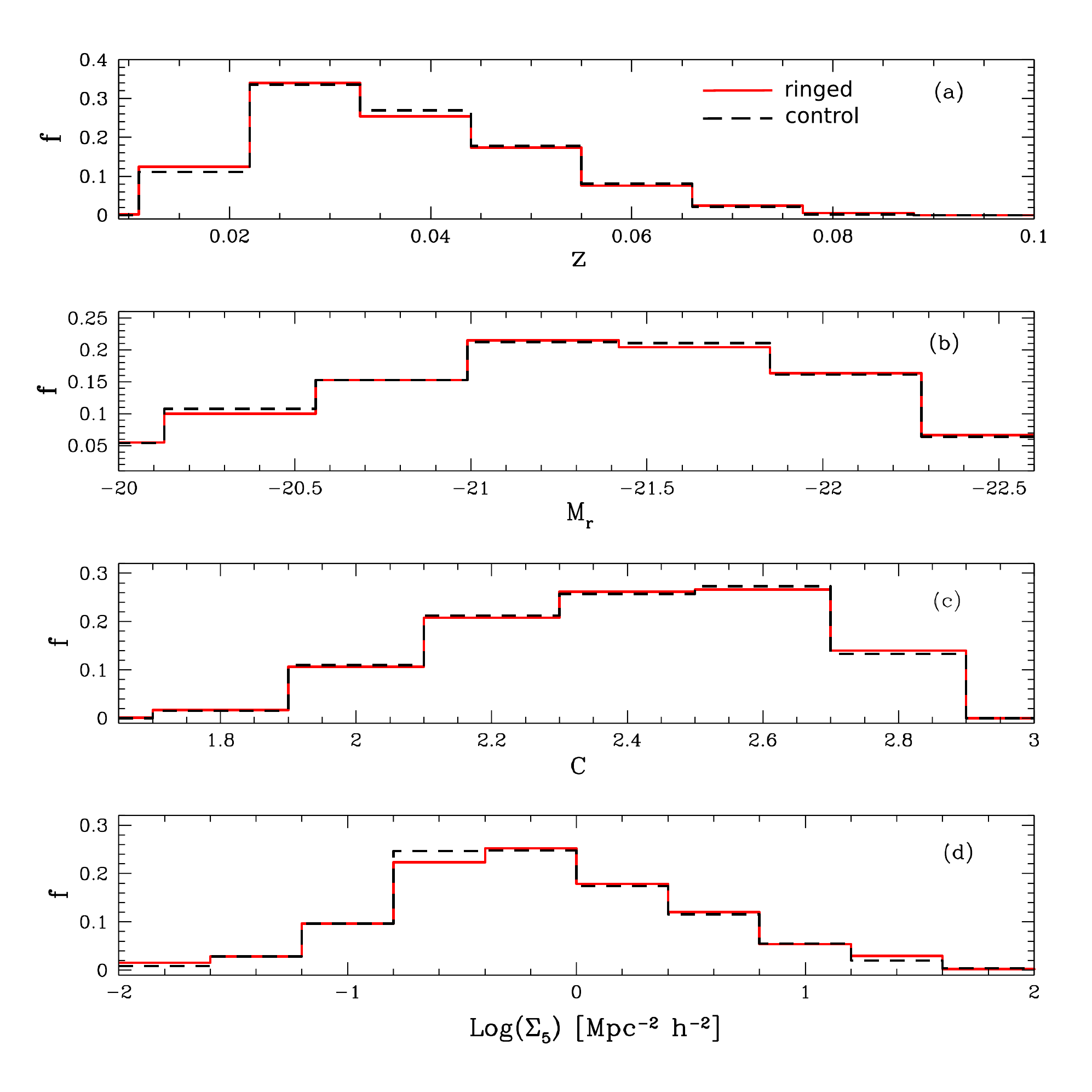} 
\caption{Normalized distributions of $z$, $\rm M_{r}$, $\rm C$, and $\rm Log (\Sigma_{5})$ for ringed galaxies (solid red lines) and for the control sample (long-dash black lines).}
\label{fig:control}
\end{figure}

With the aim to provide a reliable quantification of the effect of rings on the host galaxy properties, we built a suitable control sample taken from the 6661 non-ringed spiral galaxies (which represents 78\% of the previously classified sample), following \cite{Perez2009}.
We obtained a control catalog using a Monte Carlo algorithm that selected non-ringed
galaxies, with similar distributions of redshift and r-band absolute magnitude ($M_r$), to that of the ringed galaxy sample (see panels a and b in Fig.~\ref{fig:control}).
We also considered non-ringed galaxies in the control sample
with similar concentration index, C,  distribution to that of the
ringed catalog to obtain a similar galaxy morphology in both samples (panel c in Fig.~\ref{fig:control}).

Moreover, in order to obtain galaxies in the same local environment, we defined the projected local density parameter, $\Sigma_5$.
This parameter is calculated through the projected distance $d$ to the fifth nearest neighbor galaxy, $\Sigma_5 = 5/(\pi d^2)$ where the neighbor galaxies were chosen to have luminosities  $M_r < $-20.5  and a radial velocity difference of less than 1000 km $s^{-1}$ \citep{Balogh_2004}.  
Then, we also selected objects without rings with a similar distribution of $\Sigma_5$ to that of ringed galaxies, as shown in panel d) of  Fig.~\ref{fig:control}.

With these restrictions we obtained a control sample with the same number of objects and a similar distribution of r-band luminosities galaxy sample with ringed structures.
In addition, we applied the Kolmogorov Smirnov (KS) test to the control sample and to the ringed galaxies of the different samples, obtaining $p>$0.05 for the null hypothesis showing that the samples were drawn from the same distributions. 
Then, any difference in the galaxy features could be associated only with the presence of rings. Consequently, by comparing the results we estimated the real difference between ringed and non-ringed galaxies, unveiling the effect of these structures 
on the galaxy properties.

\subsection{Star formation activity }

The SFR represents a key parameter to understand the formation and evolution of galaxies. This is due to the fundamental relationship that exists between the mass and the local density of the gas through Schmidt's law \citep{Schmidt1959}.

 Star-forming galaxies in the low-redshift universe are mainly late-type spirals (Sb–Sdm) \citep{Kalinova2021}. These galaxies consume the gas reservoir and stop forming stars, going through a transition stage, known as the green valley, to finally become passive galaxies. This activity in galaxies can be determined through the SFR, the measurement of which, despite being a very important property, is still a challenging task. Different tracers are used to measure the SFR: recombination line emission ($\rm H_\alpha$ line), 
the UV continuum of hot stars, the far infrared radiation of thermal dust, the continuous radio emission, the CO emission from molecular clouds, among others.

In this work we use the parameter calculated by \cite{Brinchmann2004} given by $\rm Log (\rm SFR/\rm M_*) $[$\rm yr^{-1}$] (specific SFR), where $\rm M_*$ is the mass in stars. We analyzed this parameter in ringed galaxies compared to the control sample. Moreover, we studied the variation in the $\rm Log(\rm SFR/\rm M_*)$ as a function of the mass in stars and the concentration index, taking into account the different types of ringed galaxies described at the beginning of Sect. 3.
    
The upper panel of Fig.\ 4 shows the distribution of $\rm Log(\rm SFR/\rm M_*)$ for ringed galaxies and for the control sample. The distribution is bimodal where galaxies with ringed structures show a higher number with low SFRs in comparison with those from the control sample.
In the lower panels of Fig. 4 an analysis of star formation is performed for the four galaxy groups previously described. Bimodal distributions in these panels are clearly observed, each averaging close to $\rm Log(\rm SFR/\rm M_*) \sim -11 \period 3$. In addition, there are no significant differences between galaxies having internal rings and galaxies that have another types of rings. 
Moreover, from this figure it can be observed that the presence of ringed structures in a galaxy causes a decrease in the star formation with respect to galaxies without rings in the control sample.

Regarding the barred galaxies, different studies have shown that bars can induce several processes that modify many properties of the galaxies \citep{Sellwood1993, Combes1993, Zaritsky1994, Lee2012,Oh2012, Alonso2013, Alonso2014, Vera2016}. From our sample of ringed galaxies with or without bars, the distribution of the non-barred ones is shifted toward higher SFRs, compared to barred galaxies.

\begin{figure}
\centering
\includegraphics[width=0.48\textwidth]{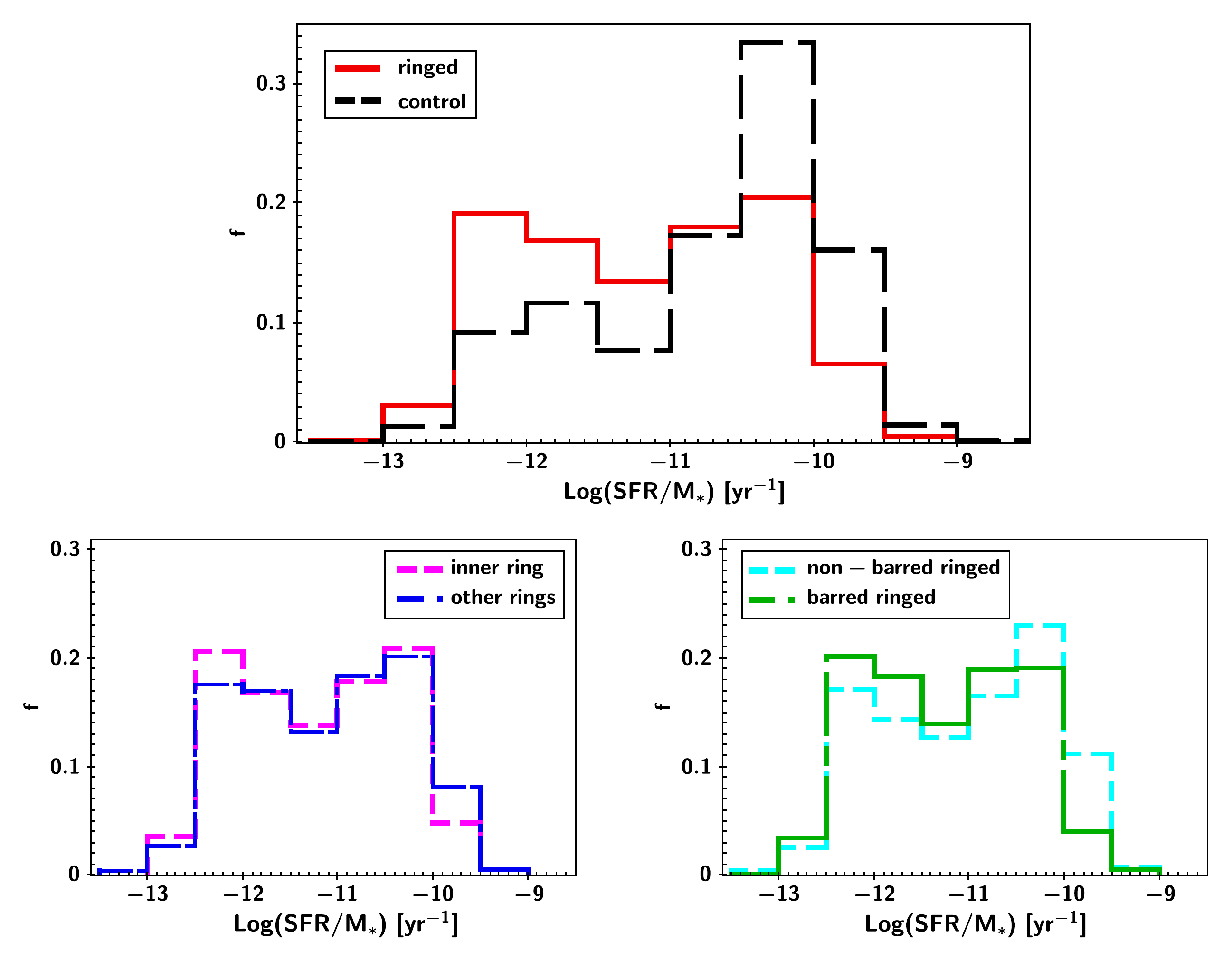} 
\caption{Specific SFR distributions for the samples considered. Upper panel: Normalized distributions of $\rm Log (\rm SFR/\rm M_*$) for ringed galaxies (solid red line) and for the control sample (long-dash black line).
Lower panels: Distributions of $\rm Log(\rm SFR/\rm M_*)$ for the different ringed galaxy types considered.
Left: Galaxies with an inner ring (dashed magenta line) and galaxies with other types of rings (two-dash blue line). Right: Non-barred ringed galaxies (dashed cyan line) and barred ringed galaxies (two-dash green line).
}
\label{fig:sfrs}
\end{figure}

\begin{figure*}
\centering
\includegraphics[width=0.45\textwidth]{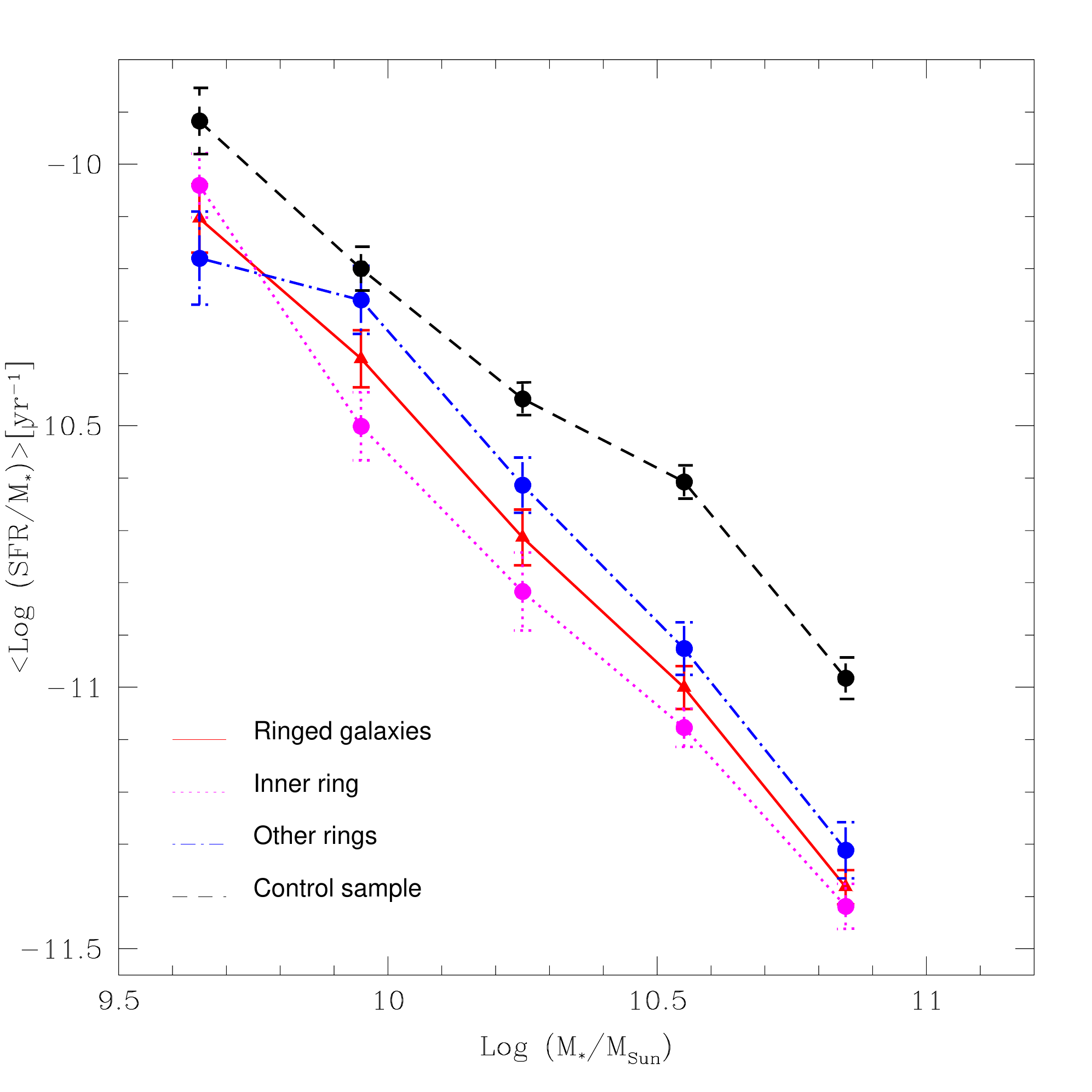}
\includegraphics[width=0.45\textwidth]{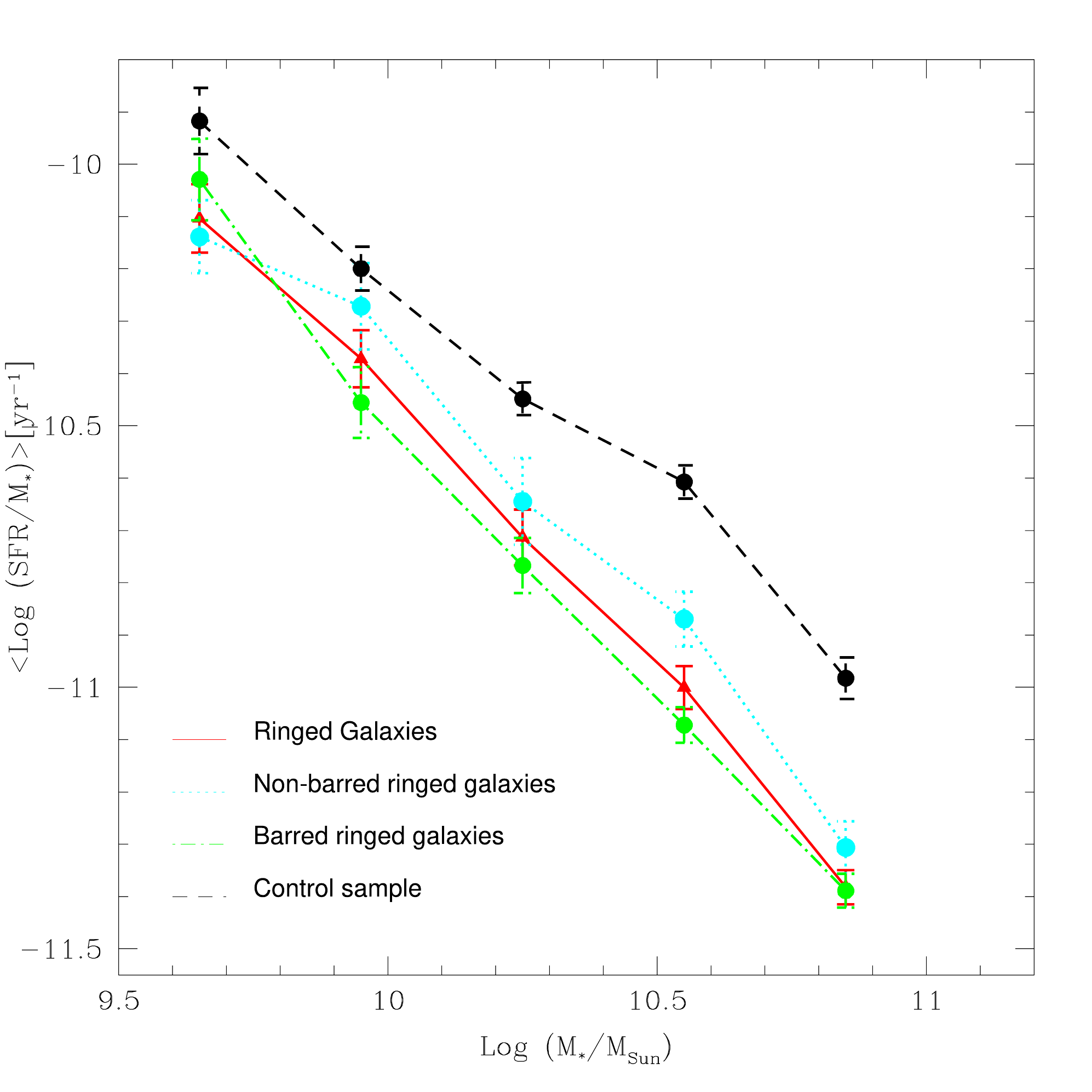}
\caption{ $< \rm Log (\rm SFR/\rm M_*) >$ as a function of Log($\rm M_*/\rm M_\odot$) for ringed galaxies (solid  red lines) and for galaxies in the control sample (long-dash black lines).
Left panel: Galaxies with an inner ring  (dashed line magenta) and galaxies with other types of rings (two-dash blue line). Right panel:  Non-barred ringed galaxies (dashed cyan line) and barred ringed galaxies (two-dash green line).
}
\label{fig:sfrmasa}
\end{figure*}

The bimodality in the $\rm Log (\rm SFR/\rm M _ *)$  distribution reflects the existence of two large groups of galaxies: the blue star-forming disk galaxies and the red ones that tend to be more inactive (with low star formation activity), mainly spheroidal or elliptical galaxies. At low $z$, passive galaxies are mainly luminous and massive, and prevail in dense environments such as groups and clusters. Conversely, star-forming galaxies exhibit lower stellar masses and are more common to find in the field \citep{Blanton2009}.
Furthermore, bimodality may be a consequence of star formation slowing down relatively quickly in some galaxies. That is the reason why a wide variety of cooling mechanisms have been proposed for the observed distributions \citep{Balogh_2004}. Thus, the SFR of star-forming galaxies strongly correlates with stellar mass, resulting in a well defined sequence.

Bearing this in mind, the relation between the mean $\rm Log(\rm SFR/\rm M_*)$ as a function of the $\rm Log(\rm M_*/\rm M_\odot)$ is presented for ringed galaxies in the four groups
and for the control sample, as shown in Fig. 5.
Errors were estimated by applying the bootstrap resampling
technique in this and in the following figures \citep{Barrow1984}.
As can be seen, star formation activity decreases toward higher stellar masses for all the samples studied in this work.
From the left panel, it can be observed that the SFR varies according to the different types of rings. It is found that galaxies with inner rings have, in general, low star formation for the entire range of masses, compared to ringed galaxies with the presence of other types of rings.
Inner, nuclear and outer rings are often lined by HII regions \citep{Buta1998} and some of them has been evolving without much further activity since the last burst of star formation \citep{Buta1996}. 
Furthermore, in the right panel of the figure, it can be appreciated that for the ringed galaxies, those with bars show lower SFRs than the non-barred ones. 
It is also observed that the slope for barred ringed galaxies is steeper than those ringed without bars, indicating that the bar could help to accelerate the star formation process \citep{Vera2016}.
Clearly, disk objects in the control sample
show more efficient activity in star formation for all stellar mass bins, with respect to the ringed galaxy samples.

\begin{figure*}
\centering
\includegraphics[width=0.45\textwidth]{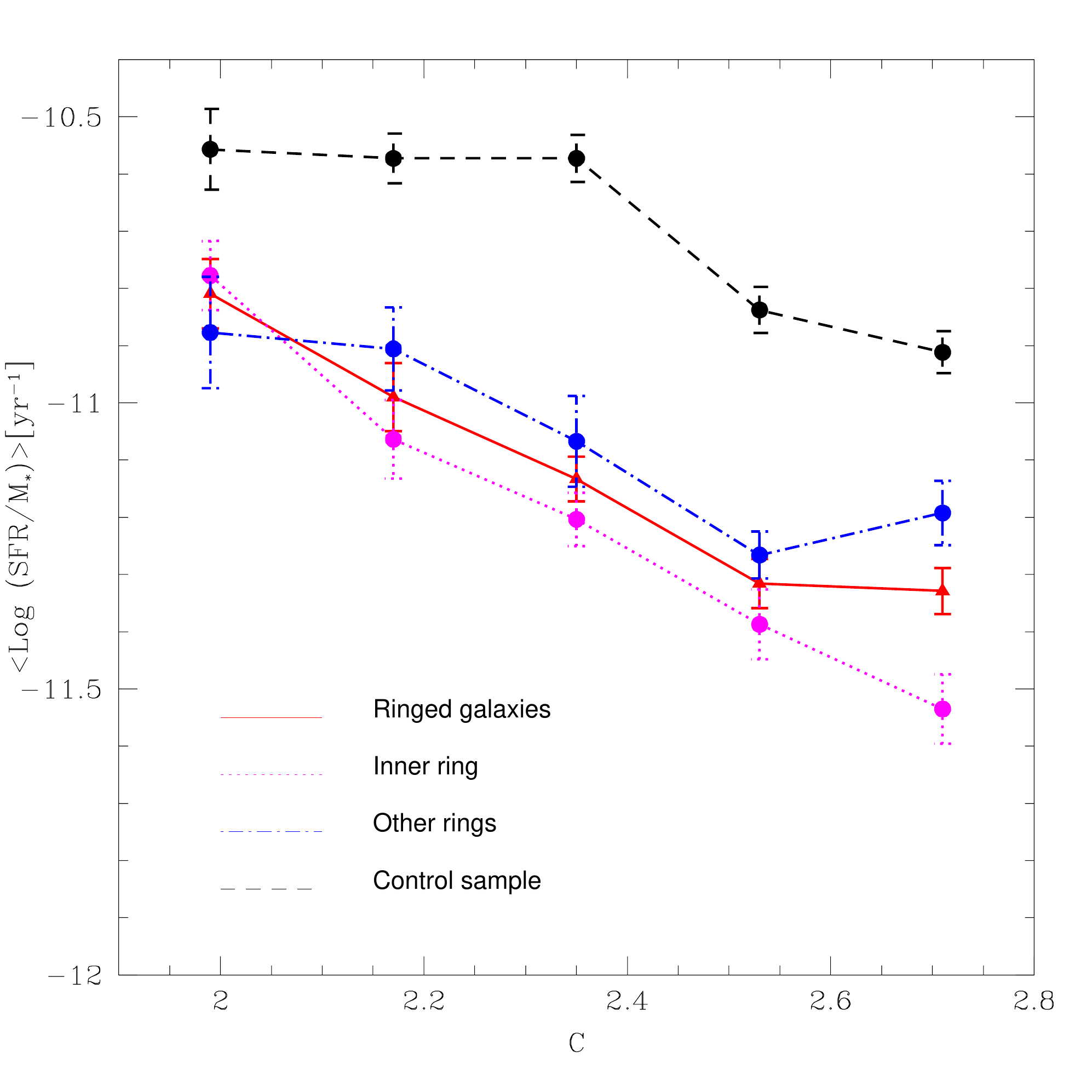}
\includegraphics[width=0.45\textwidth]{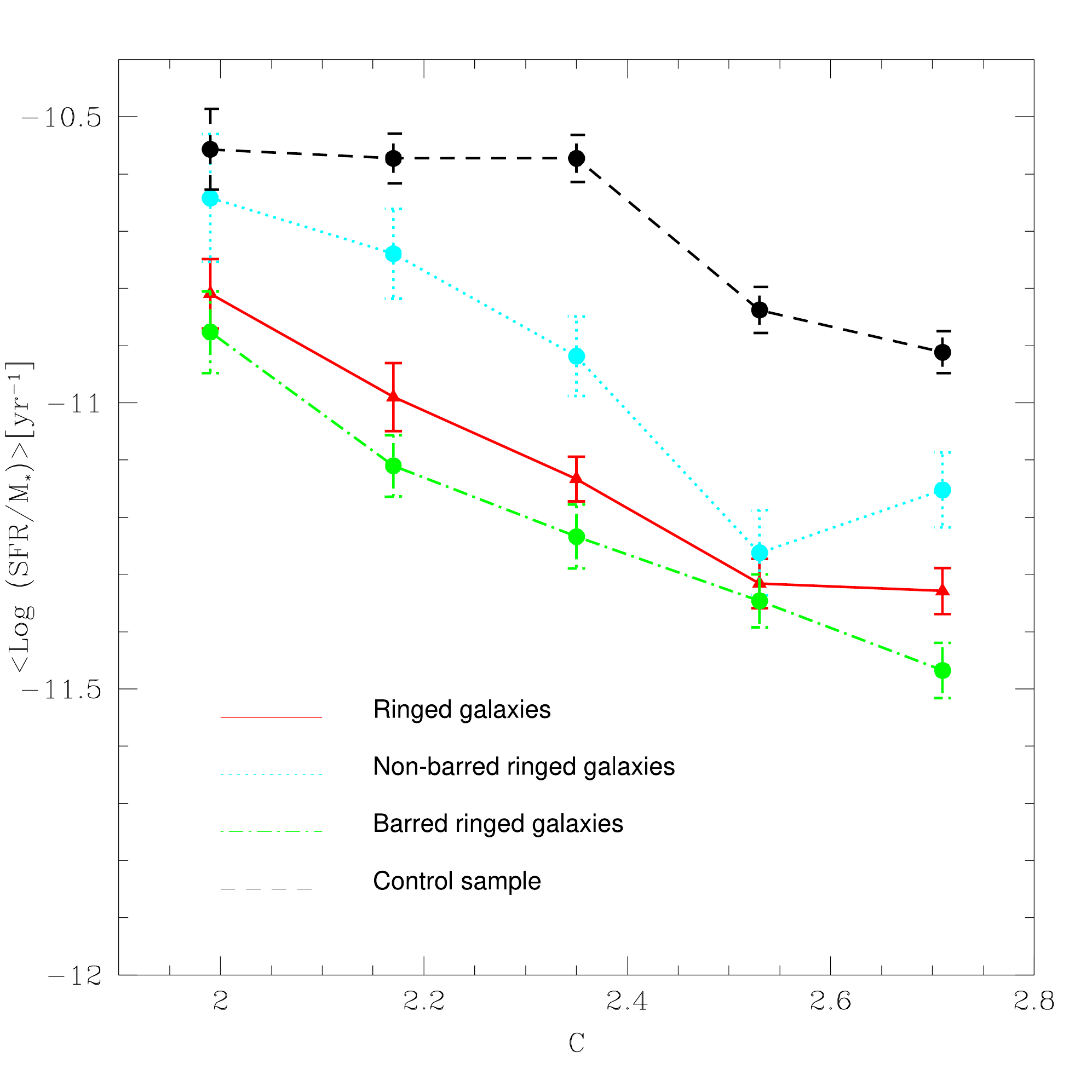}
\caption{ $< \rm Log (\rm SFR/\rm M_*) >$ versus $\rm C$ for ringed galaxies (solid red lines) and for galaxies in the control sample (long-dash black lines).
Left panel: Galaxies with an inner ring  (dashed magenta line) and galaxies with other types of rings (two-dash blue line). Right panel:  Non-barred ringed galaxies (dashed cyan line) and barred ringed galaxies (two-dash green line).
}
\label{fig:sfrc}
\end{figure*}

In addition, in Fig. 6 the $\rm Log (\rm SFR/\rm M _*)$ versus $\rm C$ parameter is shown for all the samples studied in this work. This parameter has a clear correlation with the morphological type of the galaxies.
We can clearly observe that ringed galaxies become less efficient star formers with increasing $\rm C$ index.
Moreover, for the barred and inner-ring ones the fall is more abrupt and show lower SFR values for different morphological types, than non-barred galaxies or with other types of rings, which present a similar trend to the control sample, although more pronounced.
Moreover, from this figure it can be observed that the presence of ringed structures in a galaxy causes a decrease in the star formation with respect to galaxies without rings in the control sample.
This effect is more prominent for galaxies with inner rings. Furthermore, this trend is enhanced by the presence of a bar, which seems to accelerate the stellar formation process in the host galaxies, leaving little gas, reflected in a low star formation activity.

\subsection{Stellar population}

\begin{figure}
\centering
\includegraphics[width=0.48\textwidth]{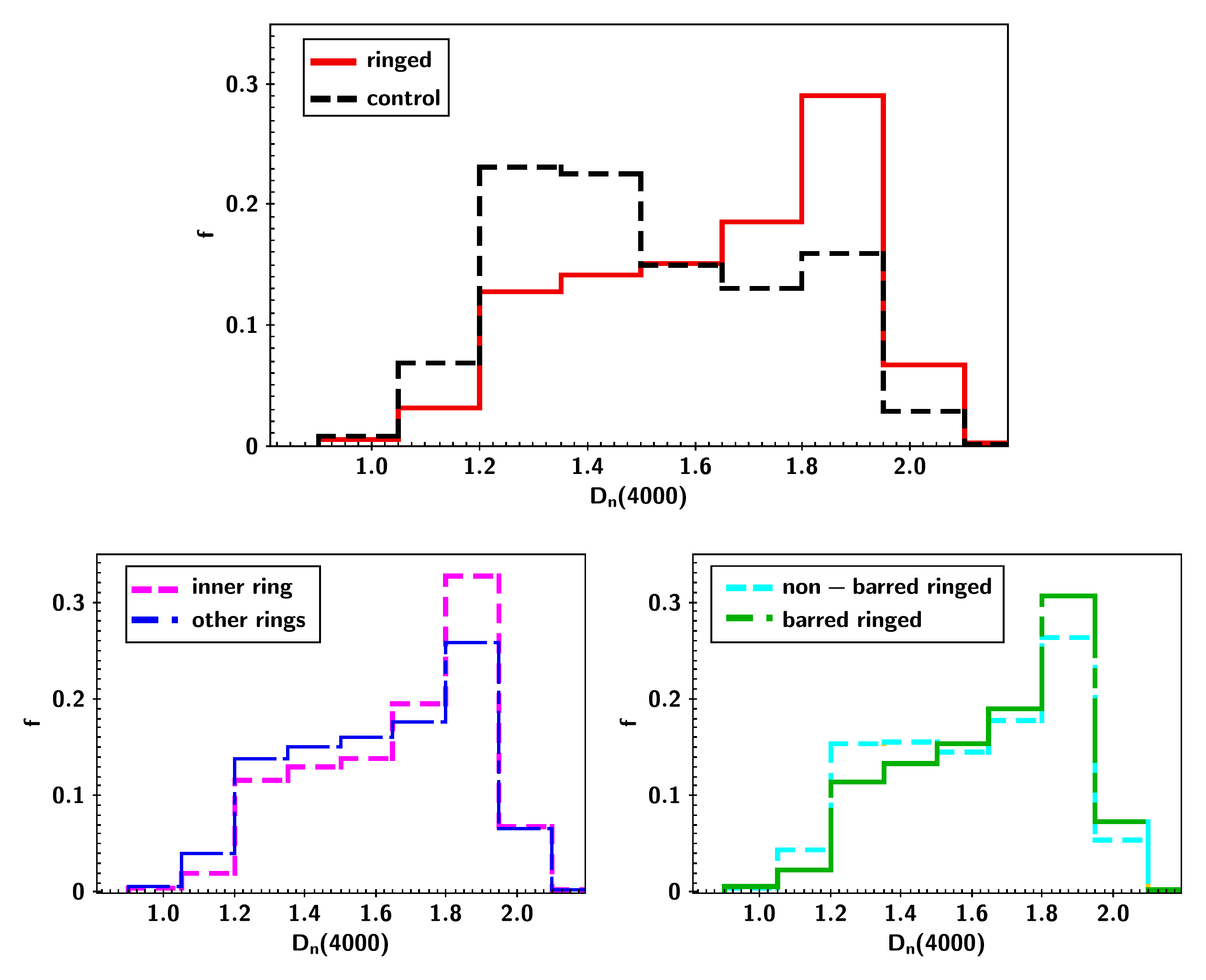} 
\caption{$\rm D_n$(4000) distributions for the samples considered. Upper  panel:  Normalized  distributions  of $\rm D_n$(4000)   for ringed galaxies (solid red line) and for the control sample (long-dash black line). Lower panels: Distributions of $\rm D_n$(4000) for the different ringed galaxy types considered. Left: Galaxies with an inner ring (dashed magenta line) and galaxies with other types of rings (two-dash blue line). Right: Non-barred ringed galaxies (dashed cyan line) and barred ringed galaxies (two-dash green line). 
}
\label{fig:dnn}
\end{figure}

\begin{figure*}
\centering
\includegraphics[width=0.45\textwidth]{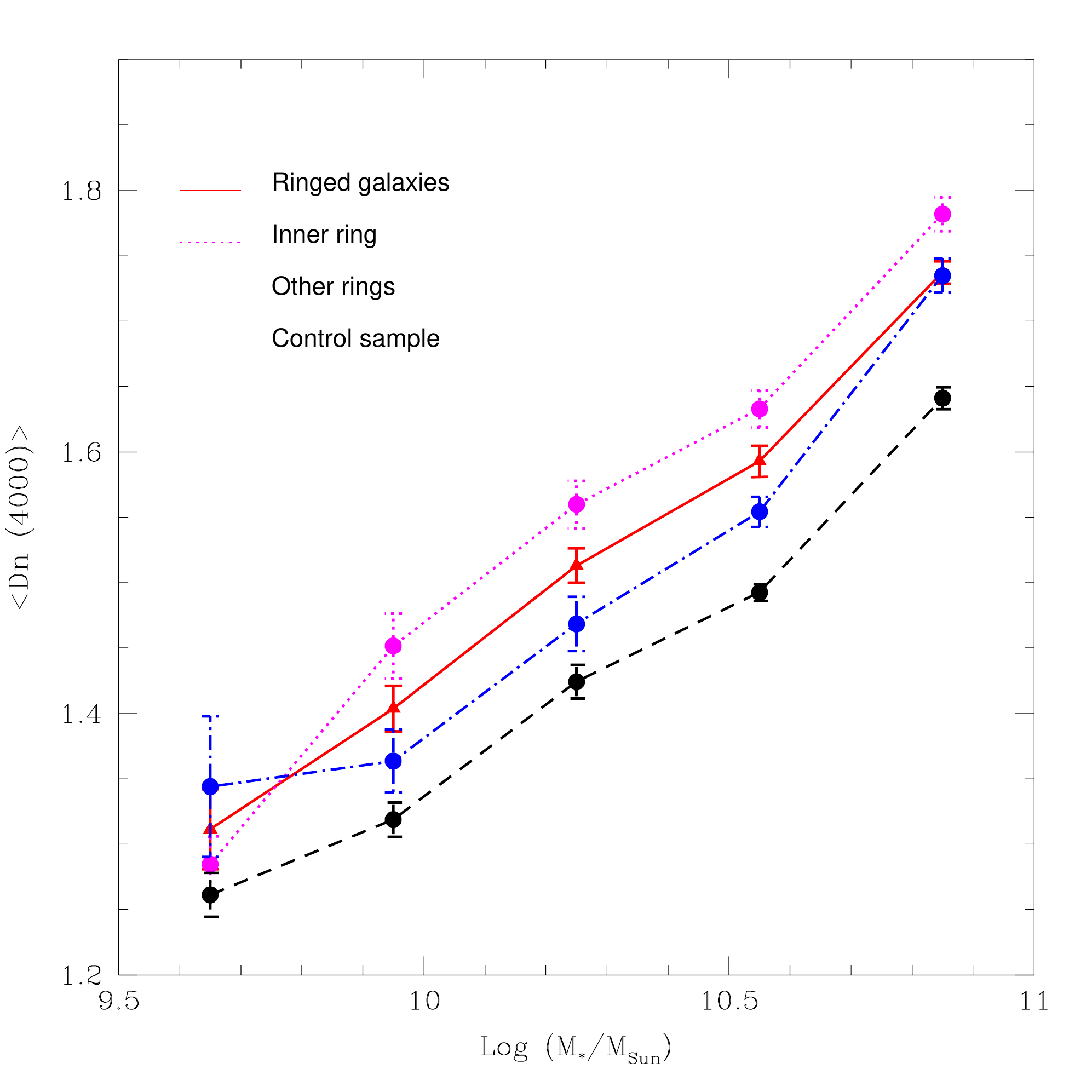}
\includegraphics[width=0.45\textwidth]{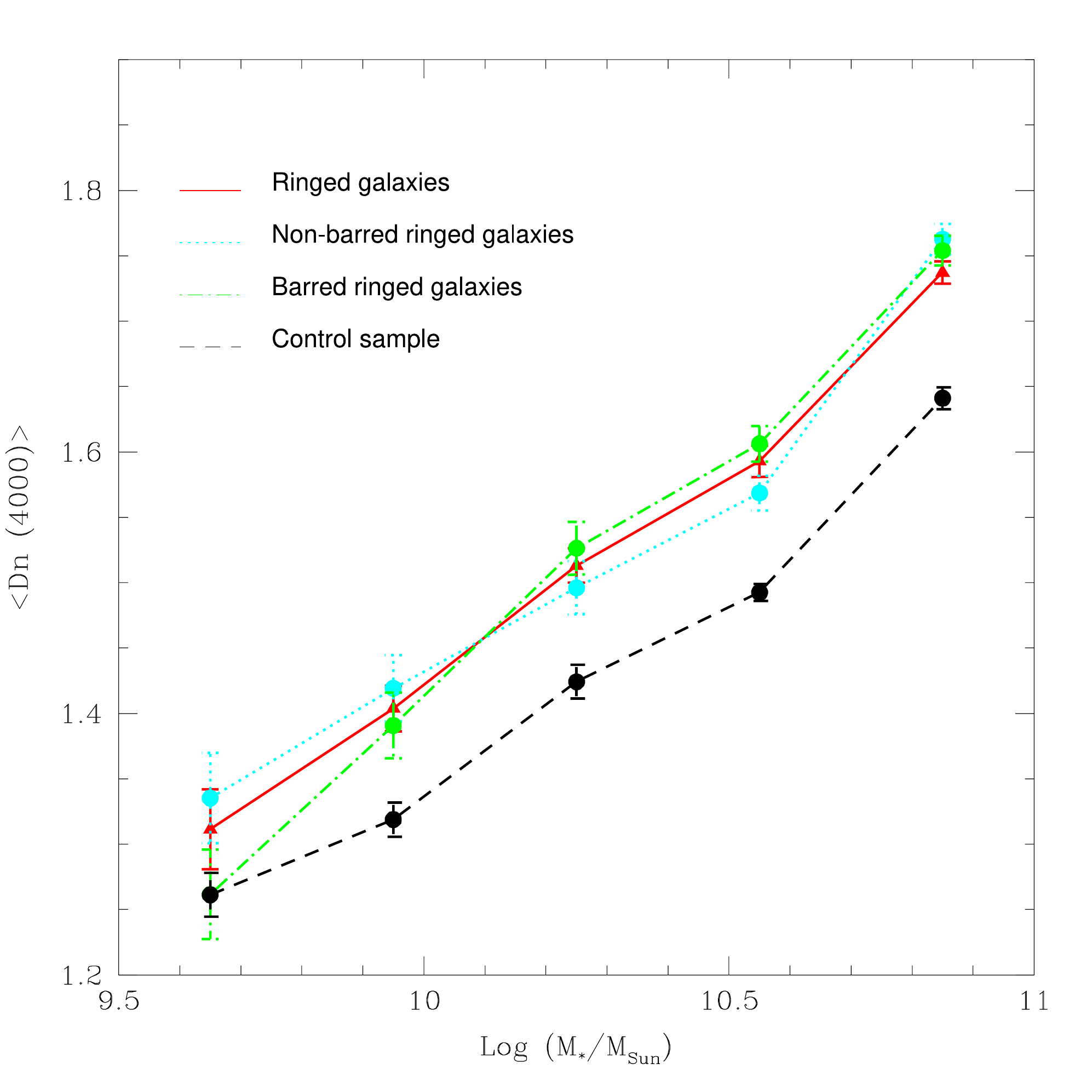}
\caption{ $< \rm D_n \rm(4000) >$ as a function of $\rm Log (\rm M_*/\rm M_\odot)$ for ringed galaxies (solid red lines) and for galaxies in the control sample (long-dash black lines).
Left panel: Galaxies with an inner ring  (dashed magenta line) and galaxies with other types of rings (two-dash blue line). Right panel: Non-barred ringed galaxies (dashed cyan line) and barred ringed galaxies (two-dash green line).
}
\label{fig:sfrmasa44}
\end{figure*}

Galaxy spectra are typically characterized by a strong break at $4000$ \AA{}, caused by the accumulation of metal absorption lines in cold stars. These lines do not exist in hot young stars because the metals are ionized. This significantly reduces the strength of this break, which makes it a good indicator of the mean age of the stellar population of galaxies.
In this work we use the parameter $\rm D_n$(4000) defined by \cite{Balogh1999}, as the ratio of the average flux densities in the 
narrow continuum bands 3850-3950 \AA{} and 4000-4100 \AA{}.

With the aim of exploring the age of the stellar population of the ringed galaxies, in Fig. 7 we show the $\rm D_n$(4000) distributions for the different types of galaxies classified previously. As shown in this figure (upper panel) the distributions for the ringed galaxies and for the control sample reach their maximums on opposite sides, being the inflection point $\rm D_n (4000)\sim 1 \period6$. Galaxies that present a ring structure have older stellar populations (higher $\rm D_n$(4000) values) than the non-ringed ones. This result is consistent with galaxies exhibiting lower star formation activity. In Fig. 7 (lower panels) different types of ringed structures are considered, showing that barred ringed galaxies and galaxies with an inner ring present older stellar populations compared to non-barred ones and with those that present other types of rings.

In addition, in Fig. 8 the $\rm D_n (4000)$ parameter as a function of $\rm Log (\rm M_*/\rm M_\odot)$ is shown for the different samples analyzed in this work. 
A clear increase in the old stellar population toward higher stellar masses is observed.
Furthermore, ringed galaxies show older stellar populations than their non-ringed counterparts throughout the whole mass range. Moreover, galaxies with internal rings exhibit a tendency toward higher $\rm Dn$(4000) values, in all $\rm Log (\rm M_*/\rm M_\odot)$ bins, than those with some other types of rings.
Regarding the barred and non-barred ringed galaxies the behavior of $\rm D_n$(4000) in both samples is similar for the entire mass range.

Figure 9 shows the relation between the parameter $\rm D_n$(4000) and the concentration index, $\rm C$. As expected, it is observed that the age of the stellar population increases toward early morphological types (higher $\rm C$ values).
Moreover, galaxies with different ringed structures present higher $\rm D_n$(4000) values compared to the control sample, independently of the $\rm C$ parameter.
We note that galaxies with an inner ring show an increase in $\rm D_n$(4000) compared to those with other types of rings, throughout concentration index range. It is also observed that this increasing trend is slightly faster and more linear for galaxies with inner rings. 
The barred galaxies show a fast growth up to $\rm C \sim 2\period17$ and then grow slowly with the morphological type, but the non-barred ones show an approximately constant growth. In addition, the barred ringed galaxies present a clear older stellar population, for all the $C$ range, with respect to their counterparts without bars.

\begin{figure*}
\centering
\includegraphics[width=0.45\textwidth]{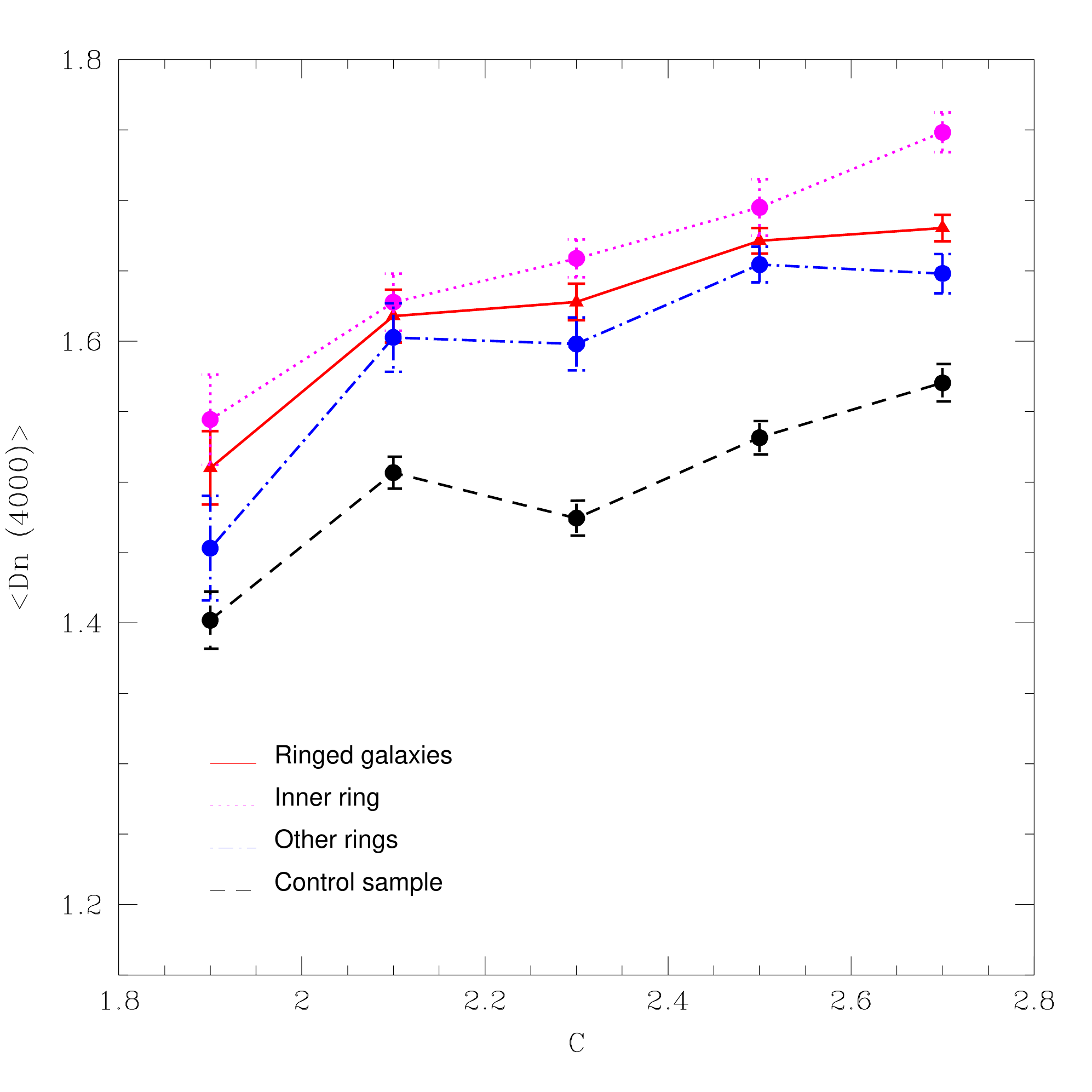}
\includegraphics[width=0.45\textwidth]{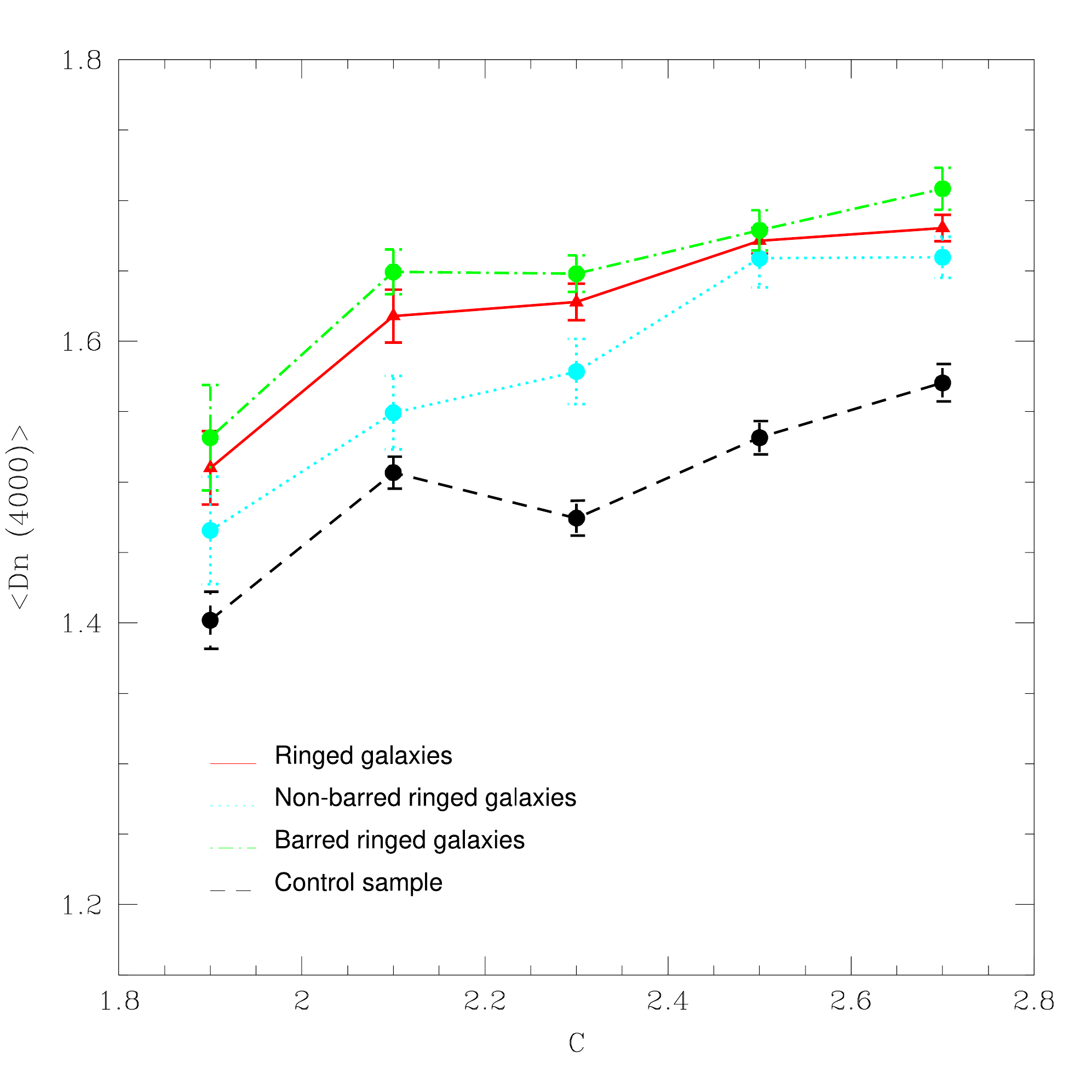}
\caption{ $< \rm D_n \rm(4000) >$ as a function of $C$ for ringed galaxies (solid red lines) and for galaxies in the control sample (long-dash black lines).
Left panel: Galaxies with an inner ring  (dashed magenta line) and galaxies with other types of rings (two-dash blue line). Right panel: Non-barred ringed galaxies (dashed cyan line) and barred ringed galaxies (two-dash green line).
}
\label{fig:sfrmasa55} 
\end{figure*}

\subsection{Galaxy colors}

With the aim of exploring the influence of ring structures on galaxy colors, in this section we assess the color indices $\rm M_u-\rm M_r$ and $\rm M_g-\rm M_r$. The importance of analyzing colors arises from the existing relationship with the dominant stellar populations and morphology.
In the local Universe, massive galaxies ($M_{*} > 10^{11}$ $\rm M_\odot$) are typically red spheroids with ancient stellar populations, which define the so-called red sequence in the color-magnitude  diagram, while the less massive (disk galaxies) have young stellar populations and reside on a more extended blue branch \citep{Strateva2001, Kauffmann2003, Baldry2004, Bernardi2003}. 

On the other hand, galaxies with intermediate colors, located in the so-called green valley are less numerous and are considered galaxies in transition from the blue branch to the red sequence, after cessation of star formation activity. A dilemma arises from the color-magnitude diagram when we try to make this distinction, since star-forming galaxies with intermediate colors between blue and red are dust obscured. On the other hand, using the colors of galaxies to describe the population, instead of morphological types, has the advantage that they are easily quantifiable and there are models that allow us to directly relate them to the history of star formation \citep{Bruzual2003}.

In Fig. 10 the distributions of both color indices ($\rm M_u-\rm M_r$ and $\rm M_g-\rm M_r$) are shown for ringed galaxies and for the control sample. We can observe that ringed objects show a clear excess of redder colors with respect to those from the control sample.
This finding is also reflected in a low efficiency in star formation activity and old stellar populations in galaxies with ringed structures, as shown in previous sections. 
Furthermore, the $ \rm M_u-\rm M_r $ distribution shows a locus that approximately separates the two peaks of each distribution at $ \rm M_u-\rm M_r \sim$ 2.2, in agreement with \cite{Strateva2001}. This value can be considered as a good limit to divide the two galaxy populations (blue and red objects).
We also observe in Fig. 11 that the color distributions have a similar behavior for galaxies with inner rings and for those with other types of ringed structures. On the other hand, when ringed galaxies have a bar there is a shift toward redder colors in the $\rm M_u-\rm M_r$ and $\rm M_g-\rm M_r$ distributions, consistent with the results observed for the previous parameters.

\begin{figure}
\centering
\includegraphics[width=0.45\textwidth]{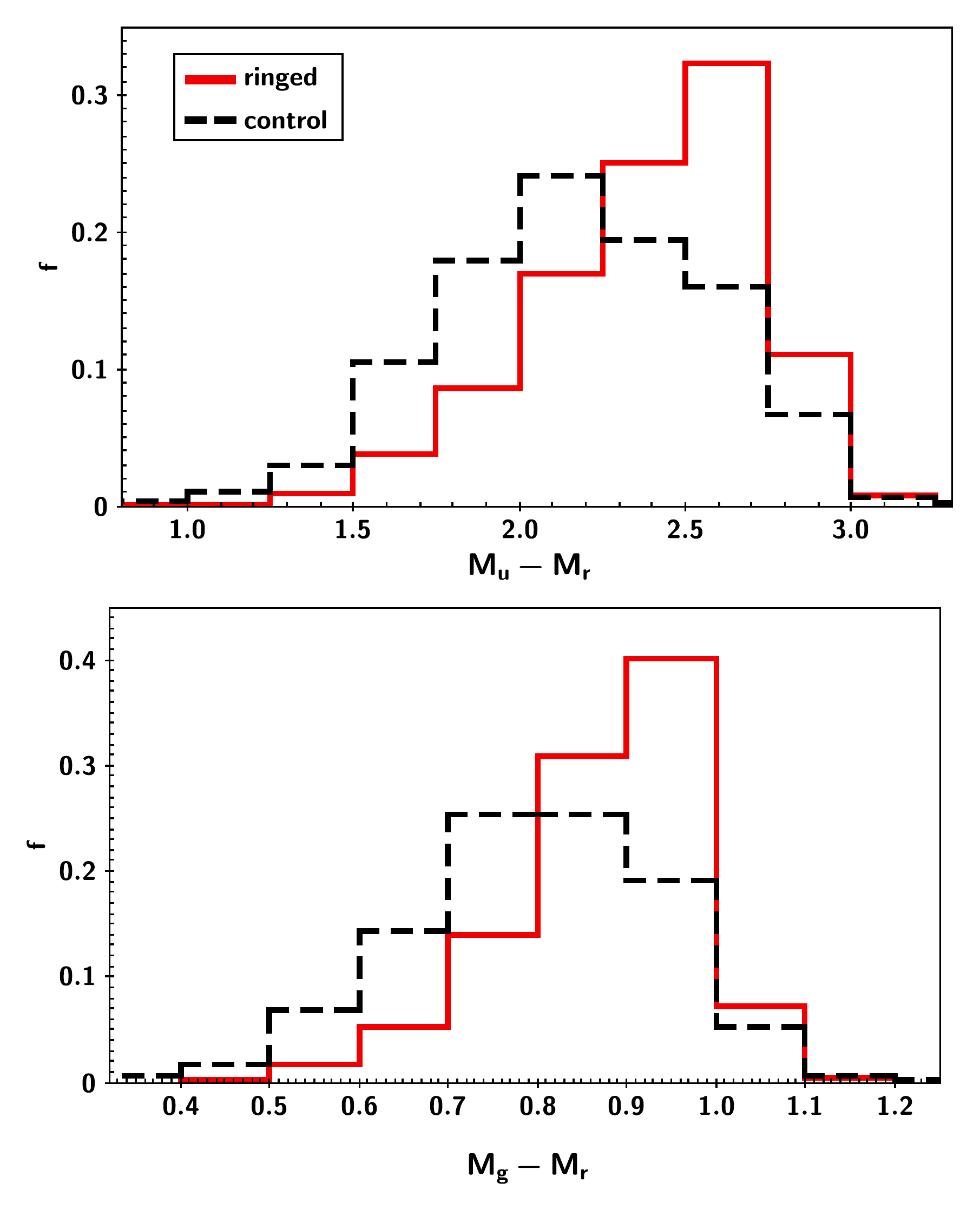} 
\caption{Normalized distributions of $\rm M_u-\rm M_r$ (upper panel) and $\rm M_g-\rm M_r$ (lower panel) for ringed galaxies (solid red lines) and for the control sample (long-dash black lines).}
\label{fig:grr}
\end{figure}

\begin{figure}
\centering
\includegraphics[width=0.48\textwidth]{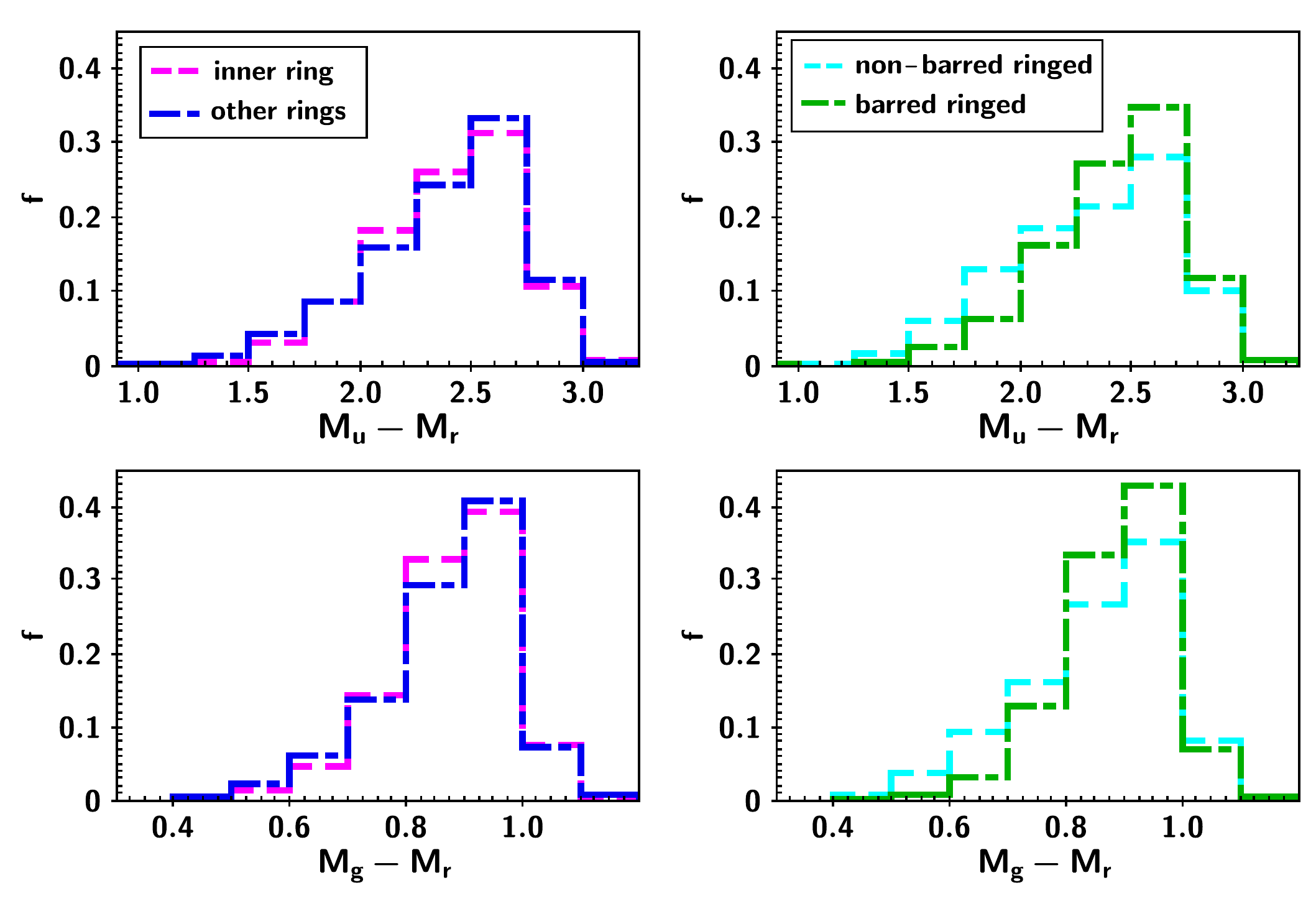} 
\caption{Color distributions for the samples considered. Left panels: $\rm M_u-\rm M_r$ and $\rm M_g-\rm M_r$ normalized distributions for galaxies with an inner ring (dashed magenta lines) and galaxies with other types of rings (two-dash blue lines). 
Right panels: $\rm M_u-\rm M_r$ and $\rm M_g-\rm M_r$ distributions for non-barred galaxies (dashed cyan lines) and barred galaxies (two-dash green lines). 
}
\label{fig:grr1}
\end{figure}

Figure 12 presents the color-magnitude diagrams for the ringed galaxies and for the control sample. 
It can be seen that ringed galaxies are principally concentrated in the top area (red region), while objects without ring of the control sample are more uniformly distributed.
We have also plotted the color fit developed by \cite{phdthesis},
which separates blue (below the line) and red (above the line) populations ($\rm M_g-\rm M_r$ = 0.15 - 0.03 $ \rm M_r $). 
As we can see, galaxies with rings are
mostly located above the line, while objects without ringed structures  lies mostly below the line (blue region).

With the aim of confirming this tendency, we considered ranges within the regions of greatest agglomeration of objects, in the $\rm M_u-\rm M_r$ diagram, to estimate which classes of galaxies are predominant in each zone. We obtained that 85 \% of ringed galaxies and 66 \% of non-ringed, respectively, fall in a color-magnitude region between $2< \rm M_u-\rm M_r<3$ and $-22<\rm M_r<-21$ (gray area in the bottom panel of the Fig. 12). 
We also divided this zone considering values of $\rm M_u-\rm M_r$ = 2.5 and $\rm M_{r}$= - 21.5, aiming to determine which galaxies are predominant and the brightest. The results show that most of the ringed galaxies fall in the upper right part of the considered area and non-ringed ones in the lower right part. Moreover, both types of galaxies fall in a range of similar brightness, suggesting that ringed galaxies are redder holding stellar populations in advanced evolutionary stages.

\begin{figure}
\centering
\includegraphics[width=0.35\textwidth]{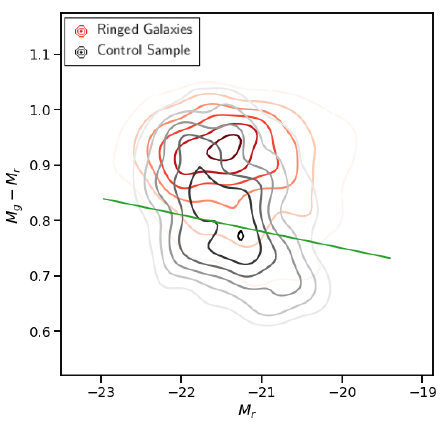} 
\includegraphics[width=0.35\textwidth]{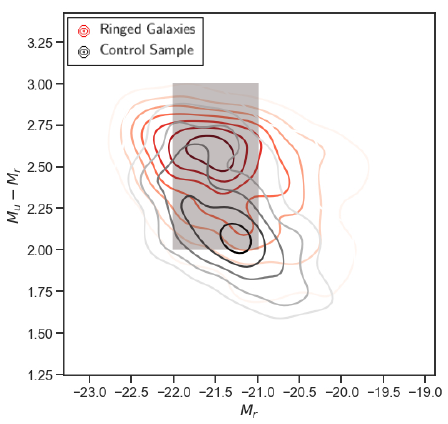} 
\caption{Color--magnitude diagrams for ringed galaxies (red outlines) and for galaxies in the control sample (black outlines). The solid green line in the upper panel represents the fit proposed by \cite{phdthesis}.
The gray area in the bottom panel represents the zone with the greatest agglomeration of galaxies. 
}
\label{fig:grr2}
\end{figure}

\begin{figure}
\centering
\includegraphics[width=0.4\textwidth]{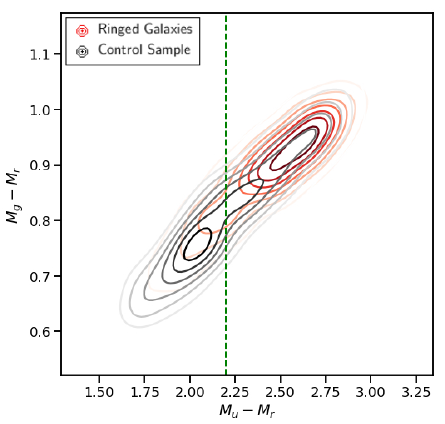} 
\caption{Color--color diagram for ringed galaxies (red outlines) and for galaxies in the control sample (black outlines). The dashed green line represents the cutoff proposed by \cite{Strateva2001}. 
}
\label{fig:grr3}
\end{figure}

Further, we analyzed the color-color diagram for galaxies with ringed structures and for the control sample, respectively. This is shown in Fig.  \ref{fig:grr3}, where a clear correlation between colors is appreciated. 
The dashed line represents the value suggested by \cite{Strateva2001} at $(\rm M_{u}-\rm M_{r})$ = 2.2, which separates the two galaxy populations,  as mentioned above. 
It can be observed that  ringed  galaxies  are  more  concentrated in the upper right part of the diagram (red sequence), while the blue cloud is more populated by galaxies without ring structures.
The plot could indicate that most ringed galaxies belong to early-intermediate spiral types, consistent with the percentages previously obtained.

This configuration could indicate evolutive states in the different classified galaxy types. 
The excess of ringed galaxies in the red sequence suggests that in  spiral galaxies with rings the drop of star formation corresponds to galaxies moving out of the blue cloud into the green valley and later to the red sequence. This happens as fast as stellar evolution allows. 
In this context, the ring structures could be a mechanism that helps to consume the gas from the disk, accelerating the star formation process. At this stage, the rings could induce many important changes in the host galaxy characteristics, producing an ascent in the color-color diagram, toward the red sequence. 
However, galaxies without ring structures show bluer colors (located at bottom left of the color-color diagram) with no signs of fast transition to the red sequence; in fact, it must take a long time for them to reach the red sequence, according to \cite{Schawinski2014}.

\subsection{Metallicity}

Metallicity is the fraction of baryonic matter that has been converted into heavier elements by means of stellar nucleosynthesis. This material may have been returned to the interstellar medium or it may still locked up in stars \citep{Kunth2000}. Therefore, stellar metallicity is a straight measurement of the amount of metals present in a galaxy, since a large part of the metals are found in its stars. 
Metallicity not only gives us an idea of the abundances of chemical elements, but also provides us with information about the fossil records of their formation history, since they are the result of several physical mechanisms acting at different stages of the galaxy evolution \citep{Freeman2002}.
The metallicity parameter used in this work is $12 + \rm Log (\rm O /\rm H)$, which represents the ratio between oxygen and hydrogen abundances \citep{Tremonti2004}. 
We found that $\approx$ 40 \% of the objects in our samples have $12 + \rm Log (\rm O /\rm H)$ measurement.

Figure 14 (top panel) illustrates the metallicity distributions for ringed galaxies and for the control sample. It can be seen that galaxies with ringed structures exhibit a higher metallicity in comparison with galaxies without rings.
We quantified the excess of disk objects with high metallicity considering galaxies above $12 + \rm Log (\rm O / \rm H)> 9\period1$, finding 64 \% for ringed galaxies and 51 \% for galaxies without rings in the control sample. 
A more detailed study of the different classes of rings (lower panels) shows that galaxies with other types of rings with respect to those with internal rings slightly extend the distribution to higher metallicities.
From this figure, it can also be seen that barred ringed galaxies present an excess toward higher metallicity values with respect to unbarred objects.

\begin{figure}
\centering
\includegraphics[width=0.48\textwidth]{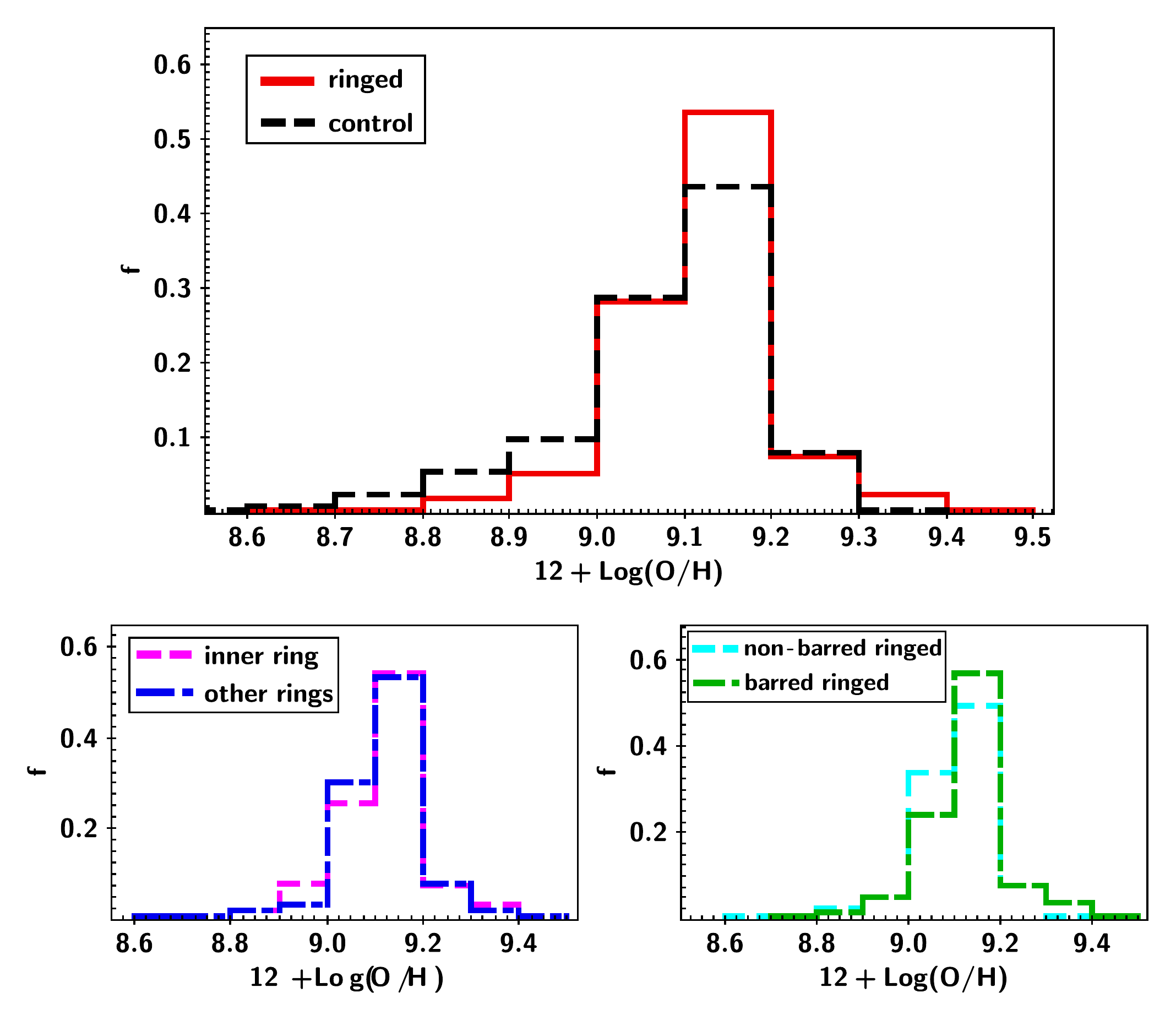} 
\caption{\label{fig:met1} 
Metallicity distributions for the samples considered. Upper panel: Normalized distributions of $12 + \rm Log (\rm O / \rm H)$ for ringed galaxies (solid red line) and for the control sample (long-dash black line). Lower left panel: Galaxies with an inner ring (dashed magenta line) and galaxies with other types of rings (two-dash blue line). Lower right panel: Non-barred ringed galaxies (dashed cyan line) and barred ringed galaxies (two-dashed green line).
}
\end{figure}

The gas metallicity is regulated by the star formation mechanism to a fixed mass, as a result of the inflow of metal-poor gas and the outflow of enriched gas \citep{Yates2012}. 
Galaxies with efficient SFR show a strong dependence on metallicity, while galaxies with low SFR show a less pronounced dependence \citep{Mannucci2010}. 
In the local Universe, \cite{Tremonti2004} studied the mass-metallicity relation (MZR; \cite{Lequeux1979}) for a sample of 53000 star-forming galaxies and confirmed the dependence of metallicity on stellar mass with high statistical significance. Moreover, \cite{Erb2006} extended this analysis to high redshift showing a similar correlation, although displaced to lower metallicity values \citep{Maiolino2007}.

To study the effect of rings on galaxy metallicity in more detail, in Fig. 15 the mass-metallicity relation for ringed galaxies and for the control sample is shown. 
It is observed that the sample of ringed galaxies presents a higher metallicity than those of the control sample.
This trend is more significant toward the lower mass ranges. 
We also compared these results with the results obtained by
\cite{Tremonti2004}, who fitted the mass-metallicity relation to a polynomial of the form $12 +\rm Log (\rm O/\rm H) = -1 \period492 + 1 \period847 (\rm Log \rm M_*)-0 \period08026 (\rm Log \rm M_*)$ for a sample of star-forming galaxies from the SDSS, finding that our galaxy samples have higher metallicity. 
These authors suggest that the mass-metallicity relation is modulated by the SFR and the metal enhancement may be due to a short-lived phase of triggered star formation in the past.
In this context, this could indicate that rings produce an accelerating effect on the gas processing and hence on the host galaxy evolution.
In the same direction, this behavior is reflected in the other galaxy features: star formation activity, stellar population and colors, showing that ringed structures produce noticeable changes in the host galaxy properties.

\begin{figure}
\centering
\includegraphics[width=0.4\textwidth]{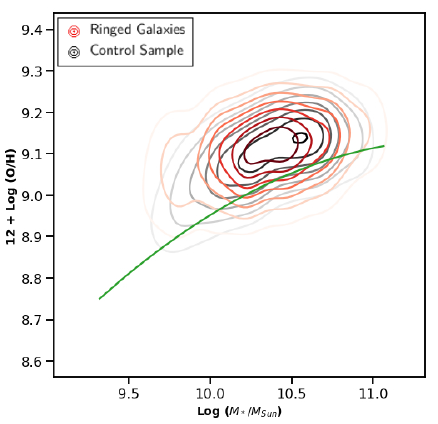} 
\caption{\label{fig:met2} Mass--metallicity relation for ringed galaxies (red outlines) and for galaxies in the control sample (black outlines). The green line represents the polynomial fit proposed by \cite{Tremonti2004}.}
\end{figure}


\section{Summary and conclusions}

We have built a catalog of galaxies with ringed structures derived from the SDSS-DR14 release, and we have performed a statistical analysis of the physical properties of ringed galaxies in contrast with non-ringed ones.
We have carried out a detailed visual classification of a sample of 8529 face-on spiral galaxies brighter than $g =$ 16.0 mag, in the redshift range 0.01 $<z<$ 0.1, based on the presence of rings and taking into account the different types of ringed structures.
In addition, we have examined morphological types and the presence of bars, lenses, and galaxy pair companions with or without interaction.
With the aim to understand how the presence of the ring affects its host galaxy's properties, we considered in particular: (i) galaxies that present an inner ring or other types of rings, (ii) barred ringed galaxies, and (iii) unbarred ringed galaxies.  
In order to provide an appropriate quantification of the effects
of rings on host galaxies, we have also constructed a suitable control
sample of non-ringed galaxies with the same redshift, r-band magnitude, concentration index, and local environment distributions.

The main results and conclusions of our analysis are summarized as follows:

(i) A sample of 1868 ringed galaxies was obtained, accounting for 22 \% of the total sample. We also found that from the total number of ringed galaxies, 46 \% have an inner ring, 10 \%  an outer ring, 20 \% both an inner and an outer ring, 6 \% a nuclear ring, and 18 \% a partial ring. 
Moreover, from the sample of galaxies with ringed structures we found that 36 \% of the objects are non-barred, 26 \% have intermediate bars, 20 \% have weak bars, and 18 \% have strong bars. We also found that 13 \% of the ringed galaxies have an internal lens and 11\% an external lens.  Regarding pairs and interactions, it was found that only 8 \% of the galaxies with ringed structures are in pair systems and that 2 \% show signs of interaction. 
 
(ii) We observed that ringed galaxies show lower star formation activity with respect to non-ringed disk objects. 
This effect is more significant for galaxies  with  inner  rings.  
These findings suggest that the presence of rings in a galaxy makes the galaxy less efficient in forming new stars and that this effect is enhanced by the presence of a bar. 
In this context, ringed structures, in conjunction with galactic bars, may slow down or stop the star formation of their host.

(iii) We also performed an analysis of the parameter $\rm D_n$(4000). We observed that it presents higher values (aged populations) in galaxies with ringed structures compared to non-ringed ones, increasing for earlier galaxy types. This could be due to the fact that, in general, ringed galaxies are at a more advanced evolutionary stage than non-ringed ones.
We also found that galaxies having inner rings and bars present higher values of $\rm D_n$(4000) than their counterparts that have some other types of rings and are non-barred objects.

(iv) From the analysis of colors, it has been found that the galaxies with ringed structures have redder populations than galaxies without rings in the control sample. This effect is more important in galaxies with inner rings and barred ringed galaxies compared to their counterparts composed of other types of rings and without bars. 
The color-magnitude and color-color diagrams show that ringed galaxies are mostly grouped in the red region, while non-ringed objects are more extended toward the blue zone. 

(v) We also explored the metallicity in ringed galaxies in comparison with non-ringed ones from the control sample. We found that galaxies with ringed structures present an excess toward higher metallicity values, while non-ringed disk objects exhibit a distribution toward lower $12 + \rm Log (\rm O / \rm H)$ values. The metallicity principally reflects the amount of gas reprocessed by the stars. In this context, it evidences that ringed structures tend to produce an increase in the metallicity of galaxies, suggesting that rings could help to process the gas contained in their hosts.

Our research suggests that galactic rings, present in a low percentage of spiral galaxies, are peculiar structures capable of strongly altering the properties of their host galaxies as well as accelerating their evolution. 
Likewise, the study of ringed galaxies still has some unresolved questions, such as regarding the ring's origin and evolution, the relation between galactic rings and the resonances or invariant manifolds, the role of the intrinsic shape in star formation in rings, etc. Therefore, the continuous investigation of these objects will allow an advance in the theoretical development proposed for their formation and evolution.\\

      This work was partially supported by the Consejo Nacional de Investigaciones Cient\'{\i}ficas y T\'ecnicas and the Secretar\'{\i}a de Ciencia y T\'ecnica de la Universidad Nacional de San Juan. V.M. also acknowledges  support  from  project  Fondecyt No. 3190736.

Funding for the SDSS has been provided by the Alfred P. Sloan
Foundation, the Participating Institutions, the National Science Foundation, the U.S. Department of Energy, the National Aeronautics and Space
Administration, the Japanese Monbukagakusho, the Max Planck Society, and the Higher Education Funding Council for England. The SDSS Web Site is http://www.sdss.org/.

The SDSS is managed by the Astrophysical Research Consortium for the Participating Institutions. The Participating Institutions are the American Museum of Natural History, Astrophysical Institute Potsdam, University of Basel, University of Cambridge, Case Western Reserve University, University of Chicago, Drexel University, Fermilab, the Institute for Advanced Study, the
Japan Participation Group, Johns Hopkins University, the Joint Institute for Nuclear Astrophysics, the Kavli Institute for Particle Astrophysics and Cosmology, the Korean Scientist Group, the Chinese Academy of Sciences (LAMOST), Los Alamos National Laboratory, the Max-Planck-Institute for Astronomy (MPIA), the Max-Planck-Institute for Astrophysics (MPA), New Mexico State University, Ohio State University, University of Pittsburgh, University of Portsmouth, Princeton University, the United States Naval Observatory, and the University of Washington.

\bibliographystyle{aa} 
\bibliography{PaperRing} 

\end{document}